\documentclass[12pt]{article}

\usepackage{a4}
\usepackage{a4wide}

\usepackage{amsmath}
\usepackage{amscd}
\usepackage{amsfonts}
\usepackage{amssymb}
\usepackage{mathrsfs}
\usepackage{fancybox} 
\usepackage{enumerate}

\usepackage{bm}
\usepackage{amscd}
\usepackage[dvips]{color}
\usepackage{graphicx}
\usepackage{subfigure}

\makeatletter

\@addtoreset{equation}{section}
\makeatother

\usepackage{color}
\definecolor{blue1}{rgb}{0.15,0.15,0.50}
\usepackage[
debug,
dvipdfm,
colorlinks=true,
urlcolor=blue1,
anchorcolor=blue,
citecolor=cyan,
filecolor=blue,
linkcolor=blue1,
menucolor=blue,
pagecolor=blue,
linktocpage=true,
pageanchor=false,
]{hyperref}    

\usepackage{amsthm} 
\newtheorem*{pf}{Proof}

\def\Tr{\mathrm{Tr}}

\def\det{\mathrm{det}}
\def\determinant#1{\underset{#1}{\mathrm{det}}}

\def\Re{\mathrm{Re}}

\def\half{{1\over2}}
\def\nn{\nonumber\\}
\def\sgn{\mathrm{sgn}}

\def\Li{\mathrm{Li}}

\def\Pint{\; {-}\hspace{-.45cm}\int}
\def\supp{\mathrm{supp}}

\newcommand{\wt}{\widetilde}

\def\={\stackrel{\bullet}{=}}

\def\({\left(}
\def\){\right)}
\def\[{\left[}
\def\]{\right]}

\def\cC{{\cal C}}
\def\cD{{\cal D}\!}

\def\cN{{\cal N}}
\def\cO{{\cal O}}
\def\cP{{\cal P}}

\def\mbf{\mathbf}

\def\mf {\mathfrak }

\def \be {\begin{equation}}
\def \ee {\end{equation}}
\def \bea {\begin{eqnarray}}
\def \eea {\end{eqnarray}}
\def \beal#1 {\begin{align}#1\end{align}}
\def \bes#1 {\begin{equation}\begin{split}#1\end{split}\end{equation}}
\def \nn {\notag\\}

\def\aver#1{\left\langle #1 \right\rangle}

\makeatletter

\@addtoreset{equation}{section}
\makeatother

\begin{document}

\begin{titlepage}
\title{
\begin{flushright}
\normalsize{ 
YITP-16-114\\
Oct 2016}
\end{flushright}
       \vspace{1.5cm}
Matrix Model of Chern-Simons Matter Theories
Beyond The Spherical Limit
       \vspace{1.5cm}
}
\author{
Shuichi Yokoyama\thanks{E-mail: shuichi.yokoyama[at]yukawa.kyoto-u.ac.jp
}
\\[25pt] 
{\it Yukawa Institute for Theoretical Physics, Kyoto University,}\\
{\it Kitashirakawa-Oiwakecho, Sakyo-Ku, Kyoto, Japan}
\\[10pt]
}

\date{}
\maketitle

\thispagestyle{empty}


\begin{abstract}
\vspace{0.3cm}
\normalsize

A class of matrix models which arises as partition function in U($N$) Chern-Simons matter theories on three sphere is investigated. 
Employing the standard technique of the $1/N$ expansion 
we solve the system beyond the planar limit. 
In particular we study a case where the matrix model potential has $1/N$ correction and give a general solution thereof up to the order of $1/N^2$. 
We confirm that the general solution correctly reproduces the past exact result of the free energy up to the order in the case of pure Chern-Simons theory. 
We also apply to the matrix model of $\cN=2$ Chern-Simons theory with arbitrary numbers of fundamental chiral multiplets and anti-fundamental ones, which does not admit the Fermi gas analysis in general.  

\end{abstract}
\end{titlepage}

\section{Introduction}
\label{intro}

Recent progress in supersymmetric Chern-Simons matter theories 
has been made on the basis of the exact result by means of the supersymmetric localization. 
This technique allows one to compute the partition function of supersymmetric theories exactly by the steepest descent method, which reduces the path integral calculation to a certain matrix model one \cite{Pestun:2007rz}. 
This calculation was done generically on $\mbf S^3$ \cite{Kapustin:2009kz,Jafferis:2010un,Hama:2010av} (see also \cite{Hama:2011ea,Imamura:2011wg}) and on $\mbf S^2\times \mbf S^1$ called superconformal index  \cite{Kim:2009wb,Imamura:2011su} (see also \cite{Bhattacharya:2008bja,Choi:2008za,Imamura:2009hc}).
These exact results were in precise agreement with the prediction from AdS$_4$/CFT$_3$ duality \cite{Drukker:2010nc,Herzog:2010hf} (see also \cite{Suyama:2009pd,Martelli:2011fu}). 
See \cite{Marino:2011nm} for review and references therein. 

On the other hand, progress in non-supersymmetric Chern-Simons matter theories has also been made not relying on the localization technique but on the $1/N$ expansion technique by restricting the class of matter fields to vector fields.  
It was conjectured that such system is exactly soluble in the 't Hooft large $N$ limit 
\cite{Giombi:2011kc,Aharony:2011jz}. 
The thermal partition function in Chern-Simons vector models on $\mbf S^2\times \mbf S^1$ was determined exactly in the leading of the $1/N$ expansion near the critical high temperature \cite{Jain:2013py} (see also \cite{Jain:2012qi,Jain:2013gza}), which was used to show the three dimensional bosonization duality \cite{Maldacena:2012sf,Aharony:2012nh,Aharony:2012ns}.

In contrast the exact large $N$ analysis of three sphere partition function for any non-supersymmetric Chern-Simons matter theory is not performed due to technical difficulty.
So far analysis of the three sphere partition function was done perturbatively near the weak coupling limit of the Chern-Simons coupling constant \cite{Klebanov:2011gs} to confirm that the system obeys the F-theorem \cite{Jafferis:2010un}. 
Perturbative analysis is, however, not enough to provide evidence for duality and exact large $N$ analysis is strongly awaited.

In this situation we change gears to study a class of matrix models which is to be obtained as the three sphere partition function of Chern-Simons matter theories in order to capture a generic feature of such a class of matrix models toward a big goal to show bosonization duality on the three sphere. 
We are also interested in analyzing such a class of matrix models beyond the planar limit because the bosonization duality is expected to hold at the subleading in the $1/N$ expansion \cite{Jain:2013gza}. (See \cite{Aharony:2015mjs,Hsin:2016blu,Radicevic:2016wqn} for recent arguments.) 

To illustrate what kind of matrix models is to be studied, let us consider the partition function of generic $U(N)_k$ Chern-Simons matter theories on the three sphere with unit radius.  
\be 
Z = \int \cD A\cD \Phi e^{-{i k\over 4\pi} \int_{\mbf S^3} ( A \wedge dA - {2i \over 3} A\wedge A \wedge A) -S[\Phi,A] },
\label{defCSMpf} 
\ee
where $\Phi$ denotes all the matter fields collectively and $S[\Phi,A]$ is the action for the matter fields. 
This may be computed perturbatively as follows. (See \cite{Adams:1997zc} for details.) 
We first expand the gauge field by the vector spherical harmonics 
\be 
A_\mu(x) = \sum_{s\in\half\mbf N}\bigg(\sum_{|l|\leq s \atop |r|\leq s+1}  a^{s}_l{}^{s+1}_{r} Y^{s}_l{}^{s+1}_{r}{}_\mu(x)+\sum_{|l|\leq s+1 \atop |r|\leq s} a^{s+1}_l{}^{s}_{r} Y^{s+1}_l{}^{s}_{r}{}_\mu(x)+\sum_{|l|,|r|\leq s} a^{s,s}_{l,r} Y^{s}_l{}^{s}_{r}{}_\mu(x) \bigg).
\ee
We take the Lorenz gauge $\nabla _\mu A^\mu=0$, which kills the modes $a^{s,s}_{l,r}$ with $s>0$. The residual gauge can kill the mode $a^{0,0}_{0,0}$ except its Cartan part, which we denote by $\sigma$. 
We expand the matter fields in a similar manner. Taking into account the Faddeev-Popov determinant we integrate out all massive modes such as $a^{s,s+1}_{l,r},a^{s+1,s}_{l,r}$ and all the modes coming from the matter fields, which are massive on the three sphere. 
Then the partition function reduces to the finite-dimensional integration of the effective action over $\sigma$.
\be 
Z = \int d^N\!\sigma e^{-i {k \over 4\pi} \sum_{s=1}^N\sigma_s^2 + \cdots },
\ee
where the ellipsis is some function of $\sigma_s$ generated by integrating out the massive modes. 
When the matter fields vanish, for example, the effective action can be exactly computed by the supersymmetric localization or the cohomological localization \cite{Kallen:2011ny} by adding the auxiliary fields so as to complete the gauge field in the $\cN=2$ vector multiplet. The result is \cite{Marino:2002fk}
\be
Z \sim \int_{\mbf R^N}\!\! d^N\!{\sigma} e^{-i {k \over 4\pi} \sum_{s=1}^N\sigma_s^2 + \sum_{t \not= s}^N \log 2\sinh({\sigma_s-\sigma_t\over2}) }.
\ee 
Then by denoting the correction of the matter fields to the effective action by  $V[\sigma]$, 
the partition function is given by a form such that 
\beal{
Z \sim&\int_{\mbf R^N}\!\! d^N\!\sigma{ \prod_{t \not= s}^N 2\sinh({\sigma_s-\sigma_t\over2}) } e^{-V[\sigma]}.  
\label{defCSV0}
}
In this note we analyze this class of matrix models by restricting the form of potential to be consisting of single trace operators: $V[\sigma]=N \sum_{s=1}^N W_\sigma(\sigma_s)$.%
\footnote{ 
This restriction corresponds to that of the representation of the matter fields which will exclude higher-dimensional representations such as the adjoint one.   
} 
The goal of this paper is to solve this class of matrix models incorporating the standard technique of $1/N$ expansion developed in the study of ordinary hermitian matrix models \cite{Brezin:1977sv} beyond the spherical limit \cite{Ambjorn:1992gw,Akemann:1996zr}. See \cite{Migdal:1984gj,Marino:2004eq} for review and further references. 

The rest of this paper is written as follows. 
In \S\ref{MM} we perform some preliminary analysis of the matrix models. 
In \S\ref{Loop} we derive the loop equation for this class of matrix models. 
In \S\ref{SolutionLoop} we solve the loop equation by using the $1/N$ expansion. 
We give a general solution for the planar limit (\S\ref{Planar}) and for the genus one (\S\ref{genus1}). Then we apply this solution to a few examples in \S\ref{application}. We first apply to pure Chern-Simons theory and compare with the known exact result to test the validity of the presented framework (\S\ref{application2pureCS}). 
We then apply to $\cN=2$ Chern-Simons theory with arbitrary numbers of fundamental and anti-fundamental chiral multiplets (\S\ref{application2N2CSV}). 
\S\ref{discussion} is devoted to discussion and future direction. 
In \S\ref{pureCS} we give a brief review on the partition function of pure Chern-Simons theory on three sphere in order for this paper to be self-contained. 

\section{Matrix model of Chern-Simons matter theories }
\label{MM}

Throughout this paper we investigate a class of matrix models such that
\beal{
Z =&\mf N\int_{\mbf R^N}\!\! d^N\!\sigma{ \prod_{t \not= s}^N 2\sinh({\sigma_s-\sigma_t\over2}) } e^{-N \sum_{s=1}^N W_\sigma(\sigma_s)},
\label{defCSV}
}
where $\mf N$ is a normalized constant and $W_\sigma(\sigma)$ is a non-singular function of $\sigma$, which can be determined at least perturbatively by integrating out the massive modes for the original Chern-Simons matter theory.
When the original theory has $\cN=2$ supersymmetry, the matrix model potential can be determined exactly by using the localization method \cite{Kapustin:2009kz}. 
For example, for $\cN=2$ $U(N)_k$ Chern-Simons theory with $N_F$ fundamental chiral multiplets with the canonical R-charge, the partition function is of a form such that \cite{Kapustin:2009kz,Jafferis:2010un}
\be 
Z^{\cN=2}_\Box =\mf N \int_{\mbf R^N}\!\! d^N\!\sigma { \prod_{t \not= s}^N 2\sinh({\sigma_s-\sigma_t\over2}) } e^{\sum_{s=1}^N -({ik \over 4\pi}\sigma_s^2 + \half i\zeta \sigma_s) + N_F \ell({-i\sigma_s \over 2\pi} +\half)}.
\label{N2CSV}
\ee
where $\ell(x)= -x \log(1 - e^{2\pi ix}) +{i \over 2} (\pi x^2+ {1\over \pi} \Li_2(e^{2\pi ix})) - {i\pi \over 12}$, and $\zeta$ is the FI parameter, which is equivalent to the real mass parameter.
Adding the same number of anti-fundamental chiral multiplets to this system, 
the partition function becomes 
\be 
Z^{\cN=2}_{\Box, \bar\Box} =\mf N\int_{\mbf R^N}\!\! d^N\!\sigma { \prod_{t \not= s}^N 2\sinh({\sigma_s-\sigma_t\over2}) } e^{\sum_{s=1}^N -({ik \over 4\pi}\sigma_s^2 + \half i\zeta \sigma_s + N_F \log \cosh {\sigma_s \over 2} )}. 
\label{N2CSQ}
\ee
See \cite{Giasemidis:2015ial,Tierz:2016zcn} for detailed analysis of this type of matrix models.

For explicit computation we write the form of potential in \eqref{defCSV} as  
\be 
W_\sigma(\sigma)= {1 \over 2\wt\lambda} \sigma^2 + \sum_{p=0}^\infty t_p e^{\sigma p}+\delta W_\sigma(\sigma)
\ee
where $\wt\lambda,t_p$ are parameters. 
In this parametrization the partition function of pure Chern-Simons theory on $\mbf S^3$, which we briefly review in \S\ref{pureCS}, is given by
\be
\wt\lambda = -2\pi i {N \over k}, \quad 
\mf N={(-)^{N(N-1)\over2} e^{-\pi (N-1)N(N+1) \over 6 i k}i^{N^2 \over 2}\over (2\pi)^N N!} , \quad t_p=0, \quad \delta W_\sigma= i{\zeta \over 2N}. 
\label{pureCSparam}
\ee
See \eqref{pureCSpfMM} for the normalization.   
The first example \eqref{N2CSV} is formally given by
\bes{ 
&{1\over \wt\lambda} = {i k \over 2\pi N} + {iN_F \over 4\pi N}, \quad t_0=-{ N_F \over N} {i\pi \over 24},\quad  t_p={i N_F\over 2N} {(-)^{p-1}\over p^2} \\  
&\delta W_\sigma(\sigma) = \sum_{p=0}^\infty u_p\sigma e^{\sigma p}, \quad u_0=i{\zeta \over 2N} - {N_F \over 4N}, \quad  
u_p= {iN_F\over 2\pi N} {(-)^{p} \over p}, 
\label{parameteru}
}
where $p\geq1$.\footnote{This expansion may not be useful for practical computation though.}
The second example \eqref{N2CSQ} is  
\be 
\wt\lambda = -2\pi i {N \over k}, \quad t_0={N_F \over N}\log 2, \quad t_p={N_F\over N}{(-)^{p} \over p} \quad (p\geq1), \quad \delta W_\sigma = {i \over 2N} \zeta + {N_F\over2N}. 
\ee

The matrix model \eqref{defCSV} can be recast in the same form as hermitian matrix models with positive eigenvalues. 
By changing integration variables so that $\phi_s=e^{\sigma_s}$, 
the partition function becomes 
\beal{
Z =&\mf N \int_{\mbf R_+^N}\!\! d^N\!\phi \prod_{t \not= s}^N (\phi_s-\phi_t) e^{-N\sum_{s=1}^N W(\phi_s)} 
\label{defCSV2}
}
where $\mbf R_+$ represents the positive real axis and 
\be
W(\phi_s)=  {1 \over 2\wt\lambda} (\log \phi_s)^2 + \log \phi_s + \sum_{p=0}^\infty t_p \phi_s^{ p}+ \delta W(\phi_s)
\label{potentialOfPhi} 
\ee
where $\delta W(\phi_s)$ is analytic on the positive real axis.
Note that the matrix model potential has the logarithmic cut on the negative real axis. 

As a result the partition function can be written by using a positive definite hermitian matrix $\Phi$ as 
\be 
Z \propto \int \cD \Phi e^{-N\Tr W(\Phi)}
\ee
with $W(\Phi)={1 \over 2\wt\lambda} (\log \Phi)^2 +\log \Phi + \sum_{p=0}^\infty t_p \Phi^{ p}+ \delta W(\Phi)$.  
This suggests that this class of matrix models can be analyzed by using the standard technique employed in ordinary hermitian matrix models.  
In what follows we show that the free energy and correlators of some sector can be determined in order in the $1/N$ expansion.

\section{Loop equation}
\label{Loop}

There is a well-known method to determine the free energy in matrix models by using the so-called resolvent, which is in the current situation defined by the the vacuum expectation value of the generating function of regular single trace operators: 
\be
\omega(z) := {1\over N}\sum_{s=1}^N\aver{ {1\over z- \phi_s}}  = {1\over N} \Tr\aver{ {1\over z- \Phi}}  ={1\over N}\sum_{p\geq0}{\aver{\Tr \Phi^p} \over z^{p+1}}. 
\ee
We remark that the resolvent is well-defined around the infinity and formally behaves as $\omega(z) \sim {1\over z}$ in the vicinity of the infinity though this asymptotic behavior is not guaranteed due to the fact that the potential \eqref{potentialOfPhi} has the logarithmic cut, which ends in the infinity. This suggests 
that the behavior of the resolvent around the infinity generally gets logarithmic correction. 
Still we can expect that there exists a limit approaching to the infinity such that the resolvent behaves as $\omega(z) \sim {1\over z}$ on a certain patch.
This request will be important to determine the resolvent later. 

Once the resolvent is determined, the coupling dependence on $\{t_p\}$ of the free energy given by $F=-\log Z$ is determined by 
\be 
{d\over dT_z} F =N^2({1 \over z}-\omega(z)) 
\label{resolvent2freeenergy}
\ee
where ${d\over dT_z}= \sum_{p\geq1}{-1 \over z^{p+1}}{\partial \over \partial t_p}$. This can be seen from the fact that ${\partial F \over \partial t_p} =N \aver{\Tr \Phi^p }$. 

In order to determine the resolvent systematically, we first derive Schwinger-Dyson equation for a generic operator $\cO[\phi]$. The vacuum expectation value of $\cO[\phi]$ is defined by 
\be 
\aver{\cO[\phi]} ={\mf N\over Z}  \int_{\mbf R_+^N}\!\! d^N\!\phi \cO[\phi] \bigg( \prod_{t \not= s}^N (\phi_s-\phi_t) e^{-N\sum_{s=1}^N W(\phi_s)} \bigg). 
\ee
Consider a one-to-one transformation on $\mbf R_+$ denoted by $\phi_s \to \phi_s'$. 
Then we obtain the identity such that
\beal{
{\mf N\over Z}  \int_{\mbf R_+^N}\!\! d^N\!\phi \cO[\phi] \bigg( \prod_{t \not= s}^N (\phi_s-\phi_t) e^{-N\sum_{s=1}^N W(\phi_s)} \bigg)
={\mf N\over Z}  \int_{\mbf R_+^N}\!\! d^N\!\phi' \cO[\phi'] \bigg( \prod_{t \not= s}^N (\phi'_s-\phi'_t) e^{-N\sum_{s=1}^N W(\phi'_s)} \bigg). \notag
}
Suppose the (infinitesimal) transformation $\phi_s'=\phi_s+ a\delta\phi_s$. 
Expanding the right-hand side in terms of $a$, we find that the zero-th order term cancels the left-hand side and the equation at the linear order gives the Schwinger-Dyson equation  
\beal{
&\aver{ \sum_{s=1}^N  ( \cO[\phi]{\partial\delta\phi_s \over \partial\phi_s} + {\partial \cO[\phi] \over \partial\phi_s} \delta\phi_s)+ 2\sum_{s>t}{\cO[\phi] \over \phi_s - \phi_t}(\delta\phi_s - \delta\phi_t) - \sum_{s=1}^N {N \cO[\phi] {\partial W(\phi_s) \over \partial \phi_s} \delta\phi_s}}  = 0. 
\label{SD}
}

To derive the loop equation from the Schwinger-Dyson equation let us choose $\cO=1, \delta \phi_s = {\phi_s \over z - \phi_s}$. Then the transformation $\phi_s\to\phi_s'$ becomes one-to-one on $\mbf R_+$, because $\delta\phi_s|_{\phi_s=0}=0$, and $a{\partial\delta\phi_s \over \partial\phi_s}= {a z \over (z - \phi_s)^2} > 0$ for $a z>0$. 
Then the left-hand side in the Schwinger-Dyson equation \eqref{SD} is computed as
\beal{
z\aver{ \sum_{s=1}^N  {1 \over (z - \phi_s)^2} + 2\sum_{s>t}{1 \over (z - \phi_s)(z - \phi_t) } - \sum_{s=1}^N N {\partial W(\phi_s) \over \partial \phi_s}{1 \over z - \phi_s}} + \aver{ \sum_{s=1}^N N{\partial W(\phi_s) \over \partial \phi_s}}. \notag
} 
The first two terms are computed as 
\beal{
& \aver{ \sum_{s=1}^N  {1 \over (z - \phi_s)^2} + 2\sum_{s>t}{1 \over (z - \phi_s)( z - \phi_t )} }
=\sum_{s,t=1}^N\aver{ {1 \over (z - \phi_s)( z - \phi_t )} }
= {d\over dT_z}  \omega(z) + N^2 \omega(z)^2. \notag
}
The third term is 
\beal{
& \aver{- N\sum_{s=1}^N {  W'(\phi_s)  \over z - \phi_s}}
=\aver{- N\sum_{s=1}^N \int_{\mbf R_+}\!\! dx \delta(x-\phi_s) {  W'(x)  \over z - x}}
=- N^2\int_{\mbf R_+}\!\! dx \rho(x) {  W'(x)  \over z - x}
\label{3rd}
}
where we define the density function  
\be 
\rho(x) := {1\over N} \Tr \aver{\delta(x - \Phi)}
= {1\over N} \sum_{s=1}^N \aver{\delta(x -\phi_s)}.  
\ee
Hereafter we assume $z$ is outside the support of the density function in order to exclude the case where \eqref{3rd} is divergent. 
The density function satisfies $\int_{\mbf R_+}\!\! dx \rho(x) = 1.$
The density function can be computed by evaluating the discontinuity of the resolvent across the real axis. 
Indeed, by using the formula 
\be 
{1\over x \mp i\epsilon}= \cP {1\over x} \pm  \pi i \delta(x) 
\label{formula}
\ee
where $x$ is a real number, $\epsilon$ is an infinitely-small positive number and $\cP$ denotes the principal value, 
the discontinuity of the resolvent between $x\pm i\epsilon$ is computed as 
\be 
\omega(x-i\epsilon) - \omega(x+ i\epsilon)= 2 \pi i \rho(x). 
\label{resolvent2density}
\ee 
By using this relation, \eqref{3rd} can be rewritten as  
\be 
\aver{- N\sum_{s=1}^N {  W'(\phi_s)  \over z - \phi_s}} 
=- N^2\oint_{\cC_{\mbf R_+}}\!\! { dw \over 2\pi i} {W'(w)  \over z - w} \omega(w)
\ee
where $\cC_{\mbf R_+}$ denotes a circle encircling $\mbf R_+$ counterclockwise.  
The 4th term vanishes because 
\beal{
\aver{ \sum_n N  W'(\phi_n)}
=&{\mf N\over Z}  \int_{\mbf R_+^N}\!\! d^N\!\phi \sum_n N  W'(\phi_n)\prod_{s\not=t} (\phi_s-\phi_t) e^{ -\sum_{s=1}^N N  W(\phi_s) } \nn
=&{\mf N\over Z} \sum_n  \int_{\mbf R_+^N}\!\! d^N\!\phi \prod_{s\not=t} (\phi_s-\phi_t) (-{\partial \over \partial \phi_n} e^{ -\sum_{s=1}^N N  W(\phi_s) }) \nn
=&{\mf N\over Z} \sum_n  \int_{\mbf R_+^N}\!\! d^N\!\phi ({\partial \over \partial \phi_n}\prod_{s\not=t} (\phi_s-\phi_t))  e^{ -\sum_{s=1}^N N  W(\phi_s) } \nn
=& \sum_n  \aver{\sum_{t\not=n} {2 \over \phi_n-\phi_t } } = 0. 
}
Collecting these we obtain the loop equation
\be
\omega(z)^2- \oint_{\cC_{\mbf R_+}}\!\! { dw \over 2\pi i} {W'(w)  \over z - w} \omega(w) +{1\over N^2}{d\over dT_z}\omega(z)= 0.  
\label{loopeq}
\ee
Note that this is of the same form as that of ordinary matrix models except the integration region.

Once the resolvent is determined by the loop equation, so are the density function by \eqref{resolvent2density} and the coupling dependence on $\{t_p\}$ of the free energy by \eqref{resolvent2freeenergy}, and similarly for other coupling dependence. 
For example, when $\delta W(\Phi) = \sum_{p\geq0} u_p \Phi^p \log\Phi$, the other coupling dependence on $\{u_p\}$ of the free energy is determined by 
\be 
{d\over dU_z} F =-N^2\upsilon(z) 
\label{freeenergyU}
\ee
where ${d\over dU_z}= \sum_{p\geq0}{-1 \over z^{p+1}}{\partial \over \partial u_p}$
and $\upsilon(z)$ is the vacuum expectation value of the generating function of singlet operators of the form $\Tr (\Phi^p\log \Phi)$: 
\beal{
\upsilon(z):=& {1\over N}\Tr \aver{\log\Phi \over z -\Phi}
= {1\over N}\sum_{s=1}^N \aver{\log\phi_s \over z -\phi_s}, 
\label{upsilon}
}
which is computed by using the resolvent $\omega(z)$ as 
\beal{
\upsilon(z)=&\int_{\mbf R_+}\!\! dx \rho(x) {\log x \over z -x}
=\oint_{\cC_{\mbf R_+}}\!\! { dw \over 2\pi i} \omega(w){\log w  \over z - w}. 
}
Similarly the $\wt\lambda$ dependence of the free energy is determined as 
\beal{
{\partial F \over \partial \wt\lambda}  =& -{\partial \log\mf N \over \partial \wt\lambda} - {N \over 2\wt\lambda^2} \aver{\sum_{s=1}^N (\log \phi_s)^2}  
= -{\partial \log\mf N \over \partial \wt\lambda}  - {N^2 \over 2\wt\lambda^2}\int_{\mbf R_+}\!\! dx \rho(x)(\log x)^2  \nn
=& -{\partial \log\mf N \over \partial \wt\lambda}  - {N^2 \over 2\wt\lambda^2}\oint_{\cC_{\mbf R_+}}\!\! { dw \over 2\pi i} \omega(w) (\log w)^2.
\label{freeenergyLambda}
}

Acting ${d \over dT_z}$ and ${d \over dU_z}$ on the free energy gives correlators of singlet operators such that  
\beal{
(\prod_{l=1}^m{d \over dU_{w_l}})(\prod_{k=1}^n{d \over dT_{z_k}})(-F)
=N^{n+m}\aver{\prod_{l=1}^m \Tr ({\log \Phi\over w_l- \Phi}) \prod_{k=1}^n \Tr ({1\over z_k- \Phi}) }_{\rm conn}
}
with $n>1$, where the subscript conn means the connected part of the correlator. 
It is also possible to compute correlators including some number of the operator $\Tr(\log\Phi)^2$ by differentiating the free energy several times with respect to $\wt\lambda$. 

We remark that 
it is not guaranteed and has to be confirmed that correlators computed this way agree with those computed from the original theory by using the path integral.%
\footnote{ 
The author would like to thank S.~Sugimoto for discussion on this point.
}
In other words, the potential of the matrix model generally depends on operators inserted in the path integral. 
This will be easily seen by considering a partition function of some supersymmetric theory computed by using the localization method. 
Since correlators of non-supersymmetric operators cannot be computed by the exact method, the potential of the matrix model cannot be reused to compute correlators of non-supersymmetric operators. In this context the matrix model potential is available when operators inserted in the path integral or parameters deforming the original theory maintain supersymmetry.

\section{Solution of the loop equation}
\label{SolutionLoop} 

In this section we solve the loop equation \eqref{loopeq} in the $1/N$ expansion. 
Analysis will depend on the large $N$ behavior of the potential. 
Generically it can be written as
\beal{ 
W(\phi)=W_0(\phi)+{1\over N}W_1(\phi), 
}
where $W_0(\phi), W_1(\phi)$ do not depend on $N$.
For explicit calculation maintaining certain extent of generality 
we study the case where the potential is given by \eqref{potentialOfPhi} with $\wt\lambda, t_p$ of order one: 
\be 
W_0(\phi) = {1 \over 2\wt\lambda} (\log \phi)^2 +\log \phi + \sum_{p=0}^\infty t_p \phi^{ p} + \cdots. 
\label{W0}
\ee
In the examples of supersymmetric Chern-Simons theory in \S\ref{MM}, this case corresponds to the 't Hooft limit with the number of flavors $N_F$ of order $N$. 
Accordingly the consistent $1/N$ expansion of the resolvent will be such that 
\beal{
\omega(z) =&  \sum_{g=0}^\infty (N^{- 2g} \omega_{g}(z)+N^{- 2g-1} \omega_{g+\half}(z))
= \sum_{\bar g\in\half \mbf N} N^{- 2\bar g} \omega_{\bar g}(z). 
}
Plugging this into the loop equation and expanding with respect to $1/N$ the loop equation is decomposed as follows. 
For $g=0$, 
\beal{
& \omega_{0}^2(z) = \oint_{\cC_{\mbf R_+}}\!\!\! { dw \over 2\pi i} {W_0'(w)\omega_{0}(w)  \over z - w}, 
\label{g0} \\
& \hat K \omega_{\half}(z) = - \oint_{\cC_{\mbf R_+}}\!\!\! { dw \over 2\pi i} {W_1'(w)\omega_{0}(w)  \over z - w},  
\label{ghalf}
}
which we call the genus zero and half loop equation respectively for convenience, 
and for $g\geq1$, 
\bes{
\hat K \omega_{g}(z) =& \sum_{g'=1}^{g-1} \omega_{g'}(z)\omega_{g-g'}(z)+\sum_{g'=0}^{g-1} \omega_{g'+\half}(z)\omega_{g-g'-\half}(z) - \oint_{\cC_{\mbf R_+}}\!\!\! { dw \over 2\pi i} {W_1'(w)\omega_{g-\half}(w)  \over z - w} + {d \over dT_z} \omega_{g-1}(z), \\
\hat K \omega_{g+\half}(z)=& 2\omega_{g}(z)\omega_{\half}(z)+  \sum_{g'=1}^{g-1}2\omega_{g'}(z)\omega_{g-g'+\half}(z) - \oint_{\cC_{\mbf R_+}}\!\!\! { dw \over 2\pi i} {W_1'(w)\omega_{g}(w)  \over z - w} +  {d \over dT_z}\omega_{g-\half}(z), 
\label{loopg}
}
where we define 
\be 
\hat K f(z) :=\oint_{\cC_{\mbf R_+}}\!\!\! { dw \over 2\pi i} {W_0'(w)  \over z - w} f(w) - 2\omega_0(z)f(z).
\ee
From these equations $\omega_{\bar g}(z)$ can be determined in order from $\bar g=0$.  
Once the resolvent is determined at the order $\bar g$ in the $1/N$ expansion, so is the density function from \eqref{resolvent2density} as 
\beal{
&\rho_{\bar g}(x)={\omega_{\bar g}(x-i\epsilon)-\omega_{\bar g}(x+i\epsilon) \over 2\pi i},
\label{resolvent2densityg}
}
where $\rho(z) =  \sum_{g=0}^\infty (N^{- 2g} \rho_{g}(z)+N^{- 2g-1} \rho_{g+\half}(z)) =  \sum_{\bar g\in \half \mbf N} N^{- 2\bar g} \rho_{\bar g}(z)$. 
The coupling dependence of the free energy on $t_p$ is determined from \eqref{resolvent2freeenergy} as
\beal{
&{d\over dT_z} F_0 = { 1\over z} - \omega_0(z), \quad  
{d\over dT_z} F_{\bar g} =- \omega_{\bar g}(z) 
\label{resolvent2freeenergyg}
}
where $\bar g\geq \half$ and $F = \sum_{g=0}^\infty( N^{2 - 2g} F_{g}+ N^{1 - 2g } F_{g+\half}) =  \sum_{\bar g\in \half \mbf N} N^{2- 2\bar g} F_{\bar g}. $

\subsection{Planar solution} 
\label{Planar}

Let us solve the planar loop equation \eqref{g0}. 
First we show that the planar loop equation contains the saddle point equation of the starting matrix model in the large $N$ limit. 
For this purpose we compute discontinuity of both sides in \eqref{g0} between $x-i\epsilon$ and $x+i\epsilon$. 
The discontinuity of the left-hand side is 
\beal{
 (\omega_0(x-i\epsilon)-\omega_0(x+i\epsilon))(\omega_0(x-i\epsilon)+\omega_0(x+i\epsilon)) 
=  2\pi i \rho_0(x) (\omega_0(x-i\epsilon)+\omega_0(x+i\epsilon)), \notag
}
where we used \eqref{resolvent2densityg}. 
That of the right-hand side is 
\beal{
\int_{\mbf R_+}\!\! { dy } W_0'(y)({ \rho_0(y) \over x -i\epsilon -y} -{ \rho_0(y) \over x + i\epsilon -y}) 
=\int_{\mbf R_+}\!\! { dy } W_0'(y) \rho_0(y) 2\pi i\delta(x-y)
= W_0'(x) 2\pi i\rho_0(x).  \notag
}
Therefore we obtain 
\be 
\omega_0(x-i\epsilon)+\omega_0(x+i\epsilon) = W_0'(x)
\label{saddlepoint} 
\ee
with $x$ in the support of the density function in the leading of the $1/N$ expansion. 
This is the same as the saddle point equation derived from the starting matrix model \eqref{defCSV2} in the large $N$ limit. 

Suppose that the support of the density function consists of $s$ distinct connected intervals, $\supp(\rho_0) =\cup_{i=1}^s [a_{2i-1},a_{2i}]$ where $0<a_1 < \cdots < a_{2s}$. 
Taking account of the fact that the loop planar equation \eqref{g0} is quadratic, 
we make each interval correspond to a square root cut of the solution.  
Under this ansatz we solve \eqref{saddlepoint}.  
Let us consider a trial function $H(z)$ which sees the deviation of the resolvent from the $s$-cut square root function $h(z)=\sqrt{\prod_{i=1}^{2s}(z-a_i)}$: 
\be 
\omega_0(z) = h(z) H(z). 
\ee
As mentioned in the previous section, we solve the loop equation so that the resolvent behaves as $\omega_0(z){\sim} {1\over z}$ in a limit approaching the infinity. 
This suggests that the trial function behaves as $H(z){\sim} {1 \over z^{s+1}}$ up to the signature and thus is analytic around the infinity. Therefore using the Cauchy theorem we find%
\footnote{ \label{differentstart}
One may more generally conclude that, for example,  
$
\oint_{\cC_\infty}\!\! {dw \over 2\pi i} { ( w^{s} + P_1(z) w^{s-1} + \cdots + P_{s-1}(z)) H(w) \over w-z } = 0, 
$
where $P_i(z)$ are polynomials of $z$. The resolvent obtained from this form in the same way as described below at first looks different from \eqref{resolvent0} but reduce to the same form by using the boundary condition, which is the same as \eqref{condition1}. 
We give a comment on this point below. 
}
\be 
\oint_{\cC_\infty}\!\! {dw \over 2\pi i} {H(w) \over w-z } = 0,
\label{starteq}
\ee
where $z$ is a complex number outside $\supp(\rho_0)$ and $\cC_\infty$ is an infinitely large circle. 
Assuming further that the trial function is analytic except the support of the leading density function, 
we can compute the left-hand side by deforming the contour to the non-analytic region:   
\beal{
&H(z) + \oint_{\cC_{\supp(\rho_0)}}\!\! {dw \over 2\pi i} {H(w) \over w - z}
= H(z) + \Pint_{\supp(\rho_0)}\!\! {dy \over 2\pi i} {W_0'(y) \over (y - z)h(y)},
} 
where we used \eqref{saddlepoint} in advance and $\cC_{\supp(\rho_0)}$ denotes a circle encircling the intervals $\supp(\rho_0)$ counterclockwise.
Therefore the trial function is determined as 
\beal{
H(z)= -\Pint_{\supp(\rho_0)}\!\! {dy \over 2\pi i} {W_0'(y) \over (y - z)h(y)}
= -\oint_{\cC_{\supp(\rho_0)}}\!\! {dw \over 2\pi i} {W_0'(w) \over (w - z)h(w)} \half, 
}
so is the planar resolvent: 
\beal{
\omega_0(z) =& {-h(z) \over 2} \oint_{\cC_{\supp(\rho_0)}}\!\! {dw \over 2\pi i} {W_0'(w) \over (w - z)h(w)}.
\label{resolvent0}
}
Then the planar density function is computed from \eqref{resolvent2densityg} as 
\be 
\rho_0(x)
= {h(x) \over \pi i} \Pint_{{\supp(\rho_0)}}\!\! {dy \over 2\pi i} {W_0'(y) \over (x - y)h(y)} 
\label{density0}
\ee
with $x \in \supp(\rho_0)=\cup_{i=1}^s [a_{2i-1},a_{2i}]$ and $h(x):= h(x-i\epsilon)$ for $x\in \mbf R_+$. 

The endpoints of the cuts are determined in the following way. 
Assume the solution obtained above behaves asymptotically as $\omega_0(z)= {1\over z}+\cdots$ approaching to the infinity.  
This is satisfied if and only if 
\beal{
\half  \oint_{\cC_{\supp(\rho_0)}}\!\! {dw \over 2\pi i} {w^k W_0'(w)\over h(w)} = \pm \delta_{k,s} 
\qquad \forall k=0, \cdots, s, 
\label{condition1}
}
where the signature is chosen suitable. 
These give $s+1$ constraints for $2s$ endpoints of the cuts, which is not sufficient unless $s=1$. 
For $s\geq2$ case, 
the residual conditions are provided by stability against the tunneling of eigenvalues between different cuts \cite{David:1990sk}. 
We demonstrate the residual condition following \cite{Jurkiewicz:1990we}.  
First we write the total matrix model potential in terms of the density function in the large $N$ limit. 
\beal{  
{V_{\rm tot} \over N^2} =& \int_{\mbf R_+}\!\!\! dx \varrho_0(x) W_0(x) - \Pint_{\mbf R_+}\!\!\! dxdy \varrho_0(x) \varrho_0(y) \log | x - y | - \mu \(\int_{\mbf R_+}\!\!\! dx \varrho_0(x) -1 \), 
\label{totalpotential}
}
where $\mu$ is Lagrange multiplier and $\varrho_0$ is the dynamical planar density function determined by the saddle point equation
\be 
W_0(x) - 2\Pint_{\mbf R_+}\!\!\! dy \varrho_0(y) \log |x - y | - \mu = 0. 
\label{saddledensity}
\ee
Differentiating this with respect to $x$ leads to \eqref{saddlepoint}, which we solved as \eqref{resolvent0}. 
This suggests that integrating \eqref{saddlepoint} with respect to $x$ does not get back to \eqref{saddledensity} because the density function is not analytic on the edges of the cuts so integration of \eqref{saddlepoint} takes different values on every intervals in general.
Requiring those values to be the same (as $\mu$) gives a nontrivial condition.%
\footnote{ 
\label{chemipot}
It is possible to consider a case where the Lagrange multiplier in \eqref{totalpotential} takes different values on each interval. In this case their values become parameters of the theory and play a role of a kind of chemical potentials. 
}
To write down the condition we define a function $\wt\mu$ on $\mbf R_+$ by   
\beal{ 
\wt\mu(x) {:=}&\Re \bigg[ W_0(x) - 2 \Pint_{\mbf R_+ }\!\! dy \wt\rho_0(y) \log (x - y) \bigg]
}  
where $\wt\rho_0(y)$ is defined by the analytic continuation of $\rho_0(y)$ from an interval to the whole positive real axis. Then the function $\wt\rho_0(x)$ takes pure imaginary values outside $\supp(\rho_0)$. 
Differentiating with respect to $x$ gives 
$\wt\mu(x)' = \Re [ - 2 \pi i \wt\rho_0(x) ]$, which suggests that the derivative of $\wt\mu(x)$ vanishes on each cut and thus $\wt\mu(x)$ is constant on each cut, as expected.
The condition for all of these constants to be equal can be written as 
$\wt\mu(a_{2i}-\epsilon) = \wt\mu(a_{2i+1}+\epsilon)$ for $\forall i=1,2, \cdots, s-1$. 
Since 
\beal{  
\wt\mu(a_{2i+1} + \epsilon) -\wt\mu(a_{2i} -\epsilon) = \int_{a_{2i}-\epsilon}^{a_{2i+1}+\epsilon} \!\!\!\!\!\! dx \wt\mu(x)'
= \int_{a_{2i}-\epsilon}^{a_{2i+1}+\epsilon}\!\!\!\!\!\! dx (- 2 \pi i \wt\rho_0(x)) 
=-\oint_{\beta_i} dw \omega_0(w) \notag
}
where $\beta_i=\cC_{[a_{2i},a_{2i+1}]}$ is a circle encircling the interval $[a_{2i},a_{2i+1}]$ counterclockwise, we obtain%
\footnote{ 
Chemical potentials mentioned in the footnote \ref{chemipot} can be added in a way that $\oint_{\beta_i} {dw } \omega_0(w) = \mu_i$ for $i=1,2, \cdots, s-1$. 
}
\be 
\int_{a_{2i}-\epsilon}^{a_{2i+1}+\epsilon}\!\!\!\!\!\! dx  \wt\rho_0(x) = 0 \qquad {\rm or} \qquad \oint_{\beta_i} {dw } \omega_0(w) = 0 \qquad 
\label{condition2} 
\ee
for $\forall i=1,2, \cdots, s-1$.
These yield the residual $s-1$ constraint equations to fix the $2s$ endpoints of the cuts. 

There is a comment on the solution \eqref{resolvent0}.
By using the condition \eqref{condition1} the solution can be rewritten in a different form such as 
\beal{
\omega_0(z) =& {-h(z) \over 2z^k} \oint_{\cC_{\supp(\rho_0)}}\!\! {dw \over 2\pi i} {w^kW_0'(w) \over (w - z)h(w)} 
\label{resolvent0another}
}
where $k=0, 1, \cdots, s$. On the other hand, this form of solution can be obtained directly by starting with a different equation from \eqref{starteq} as mentioned in the footnote \ref{differentstart}. 
Then the cuts can be determined not only by the asymptotic condition $\omega_0(z) ={1\over z}+\cdots$ with $z\sim\infty$, but also the fact that $\omega_0(z)$ is non-singular around $z\sim0$. These two conditions give the condition \eqref{condition1}. 

The free energy in the leading of the $1/N$ expansion is easily determined as  
\be 
F_0 = V_{\rm tot} - \lim_{N\to\infty}{\log \mf N\over N^2}
\label{freeenergy0}
\ee
with \eqref{saddledensity}. From this expression, we can reproduce the derivatives of the free energy with respect to the coupling constants obtained in the previous section. For example, acting ${d \over dT_z}$ on \eqref{freeenergy0} yields 
\beal{
{d F_0 \over dT_z}= \int_{\mbf R_+}\!\!\! dx \rho_0(x) {d W_0(x)  \over dT_z}
=\int_{\mbf R_+}\!\!\! dx\rho_0(x) {-x \over z(z-x)}
={1\over z} - \int_{\mbf R_+}\!\!\! dx{\rho_0(x) \over z-x}, 
}
where we used \eqref{saddledensity}, $\int_{\mbf R_+}\!\! dx \rho_0(x) = 1$ and \eqref{W0}.
This is nothing but the equation \eqref{resolvent2freeenergyg}.

\subsection{Hole correction }
\label{hole} 

The hole correction of the resolvent is determined by the genus half loop equation \eqref{ghalf}. 
Let us derive the saddle point equation at this order from \eqref{ghalf}. 
For this purpose let us rewrite \eqref{ghalf} as 
\be 
\oint_{\cC_{\mbf R_+}}\!\!\! { dw \over 2\pi i} {W_1'(w)\omega_{0}(w) + W_0'(w)\omega_{\half}(w)  \over z - w} -2 \omega_{0}(z) \omega_{\half}(z)  = 0. 
\ee
As the planar case we compute discontinuity of the left-hand side between $x-i\epsilon$ and $x+i\epsilon$. 
The discontinuity of the first term is computed as 
\beal{
\int_{\mbf R_+}\!\!\! { dy} (W_1'(y)\rho_{0}(y) + W_0'(y)\rho_{\half}(y))({1\over x  -i\epsilon- y} - {1\over x +i\epsilon - y })=
2\pi i(W_1'(x)\rho_{0}(x) + W_0'(x)\rho_{\half}(x)).  \notag
}
That of the second term is 
\beal{
-2\pi i(\rho_0(x)(\omega_\half(x-i\epsilon) + \omega_\half(x+i\epsilon)) + W_0'(x)\rho_{\half}(x)).   \notag
}
Therefore we obtain 
\be 
\omega_\half(x-i\epsilon)+\omega_\half(x+i\epsilon) = W_1'(x)
\label{saddlepointghalf} 
\ee
with $x \in \supp(\rho_0)$. 
Combining this with the planar saddle point equation \eqref{g0} we obtain 
\be 
\omega_{0,\half}(x-i\epsilon)+\omega_{0,\half}(x+i\epsilon) = W'(x)
\label{saddlepointgzerohalf} 
\ee
with $x \in \supp(\rho_0)$, 
where we set $\omega_{0,\half}:=\omega_0+N^{-1}\omega_\half$. 
This is the same form as the planar saddle point equation \eqref{g0} replacing $W_0(x)$ with $W(x)$ and the previous argument to solve this equation holds without any modification. 
Therefore {\it a solution of the usual saddle point equation is correct up to the order of hole correction!} This is a nice simplification, while the caveat is the region which \eqref{saddlepointgzerohalf} holds. That is, the region where we need to solve \eqref{saddlepointgzerohalf} is on $\supp(\rho_0)$. However when we solve \eqref{saddlepointgzerohalf} as done in \S\ref{SolutionLoop} the cut appears as the support of the density function including the hole correction, $\supp(\rho_{0,\half})$.
This small discrepancy may imply that the loop equation can be solved by assuming that {\it the support of the planar density function matches the one including the hole correction}. 
This assumption may be important to separate out the genus half one from $\omega_{0,\half}(z)$.
We expect that the discussion above will hold in more general matrix models such as two-matrix models. We leave the proof of this conjecture to future work.

\subsection{Genus one correction} 
\label{genus1} 

Let us determine the genus one correction of the resolvent. 
For simplicity we first study the case where $W_1=0$, so $\omega_\half=0$. 
In this case the genes one loop equation, \eqref{loopg} with $g=1$, reduces to 
\be 
\hat K \omega_1(z) = {d \over dT_z} \omega_0(z). 
\ee
This can be solved in the same manner as in the ordinary hermitian matrix model \cite{Ambjorn:1992gw,Akemann:1996zr}. 
Let us first compute the right-hand side. 
\beal{
{d \omega_{0}(z)\over dT_z}=&  {d\log h(z) \over dT_z} \omega_0(z) +{h(z) \over 2} \oint_{\cC_{\supp(\rho_0)} }\!\! {dw \over 2\pi i} {  {d W_0'(w) \over dT_z} - W_0'(w) {d \log h(w) \over dT_z} \over (z -w) h(w)} \nn
=&{d\log h(z) \over dT_z} \omega_0(z) + {h(z) \over 2} \oint_{\cC_{\supp(\rho_0)} }\!\! {dw \over 2\pi i} { -1 \over (z -w)^3 h(w)} +{h(z) \over 2} \oint_{\cC_{\supp(\rho_0)}}\!\! {dw \over 2\pi i} { - W_0'(w) {d \log h(w) \over dT_z} \over (z -w) h(w)}. \notag
} 
Then the second term is computed as 
\beal{
{h(z) \over 2} \oint_{\cC_{z} }\!\! {dw \over 2\pi i} { 1 \over (z -w)^3 h(w)}
= - {h(z) \over 2} \half ({ 1 \over h(z)})''
={-1\over 4}({3\over 4} \sum_{i=1}^{2s} {1\over (z-a_i)^2} + {1\over 2} \sum_{i<j} {1\over (z -a_i)(z-a_j) }  ). \notag
}
The third term is 
\beal{
{h(z) \over 2} \sum_{i=1}^{2s} {d (-a_i) \over dT_z} \oint_{\cC_{\supp(\rho_0)}} {dw \over 2\pi i} { - W_0'(w) \over 2(z -w) h(w) (w-a_i)}
=-{d\log h(z) \over dT_z} \omega_0(z) + \sum_{i=1}^{2s} {d a_i \over dT_z}{1\over z-a_i}  {1\over 4} h(z)  M_i^{(1)},   \notag
}
where we set 
\be
M_i^{(k)} :=  \oint_{\cC_{\supp(\rho_0)}}\!\!{dw \over 2\pi i} { W_0'(w) \over h(w)(w - a_i)^k}.
\label{Mik} 
\ee
Therefore we obtain 
\be
{d \omega_{0}(z)\over dT_z}= -({3\over 16} \sum_{i=1}^{2s} {1\over (z-a_i)^2} + {1\over 8} \sum_{i<j} {1\over (z -a_i)(z-a_j) }  ) + \sum_{i=1}^{2s} {d a_i \over dT_z}{1\over z-a_i}  {1\over 4} h(z)  M_i^{(1)}. 
\label{domegadT1}
\ee
In order to compute ${d a_i \over dT_z}$, we act ${d\over dT_z}$ on the constraint equations of the edges of the cuts, \eqref{condition1}, \eqref{condition2}. 
\beal{ 
& \oint_{\cC_{\supp(\rho_0)}}\!\!{dw \over 2\pi i} w^k{ {d \over dT_z} W_0'(w) - W_0'(w){d \over dT_z} \log h(w) \over h(w)} 
= 0, \qquad k=0, 1, \cdots, s, \\
& \int_{a_{2l}}^{a_{2l+1}}\!\!\! dx {d \over dT_z}(W_0'(x) - 2\omega_0(x) )= 0, \qquad  l=1,2, \cdots, s-1
}
where we used $2\pi i \wt\rho_0(x)= W_0'(x) - 2\omega_0(x)$. These are computed as 
\beal{
&{k z^{k-1} \over h(z) } + \half \sum_{i=1}^{2s} ( a_i^k {d a_i \over dT_z} M_i^{(1)} - {z^k\over h(z)(z-a_i)} ) = 0, \\
&\half\sum_{i=1}^{2s} {K_{l,i} \over h(z)(z - a_i)}  + {1\over 2} \sum_{i=1}^{2s} K_{l,i} {d  (-a_i) \over dT_z} M_i^{(1)}  = 0,
}
where $K_{l,i} := \int_{a_{2l}}^{a_{2l+1}}dx { h(x) \over (x- a_i) }$. 
The solution can be written as 
\be
{d  a_i \over dT_z} ={1\over M_i^{(1)}}( {1 \over h(z)(z - a_i)} + \sum_{l'=0}^{s-2}\alpha_{i,l'} {z^{l'} \over h(z) }) 
\label{dadT}
\ee
where $\alpha_{i,l}$ are determined by plugging this back and setting the coefficients of polynomials with respect to $z$ to zero. 
The determining equations of $\alpha_{i,l}$ are 
\beal{
&\sum_{i=1}^{2s} ( a_i^k \alpha_{i,l'} - a_i^{k-1 -l'} )= 0,  \quad (0\leq l' \leq k-2), \quad  
\half \sum_{i=1}^{2s} a_i^k \alpha_{i,k-1}  + k-\half = 0, \\
&\sum_{i=1}^{2s}  a_i^k \alpha_{i,l'} = 0,  \quad (k\leq l' \leq s-2), \\
&\sum_{i=1}^{2s} K_{l,i} \alpha_{i,l'}{=} 0, \quad (1\leq l \leq s-1, \; 0 \leq l'\leq s-2).
}
Substituting \eqref{dadT} into \eqref{domegadT1} we find 
\beal{
{d \omega_{0}(z)\over dT_z}={1\over 16} \sum_i {1\over (z-a_i)^2} - {1\over 8} \sum_{i<j}{1\over (z -a_i) (z-a_j) } + {1\over 4} \sum_{i=1}^{2s} \sum_{{l'}=0}^{s-2} {1\over z-a_i} \alpha_{i,{l'}} a_i^{l'}. 
\label{domegadT2}
}
This can be rewritten as the image of the linear operator $\hat K$ in such a way that 
\beal{
{d \omega_{0}(z)\over dT_z}=\hat K \bigg [{1\over 16} \sum_i \chi_i^{(2)}(z) - {1\over 8} \sum_{i<j}{\chi_i^{(1)}(z) -\chi_j^{(1)}(z) \over a_i -a_j}+ {1\over 4} \sum_{i=1}^{2s} \sum_{{l'}=0}^{s-2} \chi_i^{(1)}(z) \alpha_{i,{l'}} a_i^{l'} \bigg], 
}
where $\chi_i^{(n)}(z)$ satisfies $\hat K \chi_i^{(n)}(z) = {1\over (z-a_i)^n}$ for $n \geq1$.   
$\chi_i^{(n)}(z)$ is constructed inductively as follows. 
Start with the identity such that 
 \beal{  
 {1\over (z-w)(w-a_i)^\mf n} = {1\over (z-w) (z-a_i)^\mf n}+ \sum_{k=1}^{\mf n} {1\over (z-a_i)^{k} (w-a_i)^{\mf n-k+1}} 
 }
for $ \forall \mf n \geq 1$. 
Acting $ \oint_{\cC_{\supp(\rho_0)} }\!\!\!  {dw \over 2\pi i} {W_0'(w) \over h(w)}$ on both sides  and computing the right-hand side results in 
\beal{
\!\!\!\!\!\!\!\!\!\! 
\!\!\!\!\!\!\!\!\!\! 
\int_{\cC_{\supp(\rho_0)} } \!\!\! {dw \over 2\pi i} { W_0'(w) \over h(w)}{1\over (z-w)(w-a_i)^\mf n} =&  2 \omega_0(z) {1\over h(z)(z-a_i)^\mf n} + \sum_{k=1}^{\mf n}  {M_i^{(\mf n-k+1)}\over (z-a_i)^{k} }. 
}
Equivalently, 
\be 
\hat K ({1\over h(z)(z-a_i)^\mf n})=\sum_{k=1}^{\mf n}  {M_i^{(\mf n-k+1)}\over (z-a_i)^{k} }.
\qquad
\label{lemma1}
\ee
The case with $\mf n=1$ implies $\chi_i^{(1)}(z)=  {1\over M_i^{(1)}} {1\over h(z)(z-a_i)}$. 
Assuming $\chi_i^{(n)}(z)$ is constructed so that $ \hat K \chi_i^{(n)}(z) = {1\over (z-a_i)^n}$ is valid for $n \leq\mf n-1$, we can rewrite the right-hand side as 
${M_i^{(1)}\over (z-a_i)^{\mf n} }+ \sum_{k=1}^{\mf n-1} M_i^{(\mf n-k+1)}  \hat K \chi_i^{(k)}(z)$. Therefore we find 
\be 
\hat K \({1\over M_i^{(1)}} \( {1\over h(z)(z-a_i)^\mf n}-  \sum_{k=1}^{\mf n-1} M_i^{(\mf n-k+1)} \chi_i^{(k)}(z) \) \)={1\over (z-a_i)^{\mf n} }. 
\ee
Hence if we define $\chi_i^{(\mf n)}(z)$ by 
\be 
\chi_i^{(\mf n)}(z) = {1\over M_i^{(1)}} \( {1\over h(z)(z-a_i)^\mf n}-  \sum_{k=1}^{\mf n-1} M_i^{(\mf n-k+1)} \chi_i^{(k)}(z) \), 
\label{chi}
\ee
then $ \hat K \chi_i^{(n)}(z) = {1\over (z-a_i)^n}$ is valid for $n = \mf n$. 
By using this function we finally solve the genus one loop equation as 
\be
\omega_1(z) ={1\over 16} \sum_{i=1}^{2s} \chi_i^{(2)}(z) - {1\over 8} \sum_{i<j}{\chi_i^{(1)}(z) -\chi_j^{(1)}(z) \over a_i -a_j}+ {1\over 4} \sum_{i=1}^{2s} \sum_{{l'}=0}^{s-2} \chi_i^{(1)}(z) \alpha_{i,{l'}} a_i^{l'}
\label{resolvent0g1}
\ee
up to terms in the kernel of the operator $\hat K$ such as ${z^m \over h(z)}$ with $m=0, \cdots, s$. 
Remark that $\omega_1(z)$ behaves at most ${1\over z^{s+1}}$ and thus the leading asymptotic behavior of the total resolvent $\omega(z)$ is unchanged, so is the cut.     
From \eqref{resolvent0g1} the coupling dependence of the genus one free energy can be determined. 
For example the dependence on $t_p$ is determined by \eqref{resolvent2freeenergyg} and that on $\wt\lambda$ is by \eqref{freeenergyLambda}.%
\footnote{ 
The differential equation \eqref{resolvent2freeenergyg} will be solved as done in the original hermitian matrix model for one cut case \cite{Ambjorn:1992gw} and two cut case \cite{Akemann:1996zr}, though such explicit solutions of the genus one free energy do not contain the information of the dependence on other coupling constants such as $\wt\lambda$.  
}

Next we consider the case where the matrix model potential contains $1/N$ correction: $W_1(w)\not=0$. 
In this case, as derived in \eqref{loopg}, the genus one loop equation is corrected by the genus half resolvent so that $\hat K \omega_{1}(z) = \omega_{\half}(z)^2 - \oint_{\cC_{\mbf R_+}}\!\!\! { dw \over 2\pi i} {W_1'(w)\omega_{\half}(w)  \over z - w} + {d \over dT_z} \omega_{0}(z).
$
This equation is more involved and there may be simplification in a way that the planar resolvent and the genus half one can be determined at the same time as shown in \S\ref{hole}. 
To see this let consider the deviation of the resolvent from the solution $
\omega(z) = \omega_{0,\half}(z)+\delta\omega(z)$
and substitute this into the original loop equation \eqref{loopeq}. 
We obtain 
\beal{
2\omega_{0,\half}(z)\delta\omega(z) +\delta\omega(z)^2 - \oint_{\cC_{\mbf R_+}}\!\!\! { dw \over 2\pi i} {W'(w)  \over z - w} \delta\omega(w) +{1\over N^2}  {d \over dT_z}( \omega_{0,\half}(z) +\delta\omega(z) ) = 0.
}
Since $\omega_{0,\half}$ is of the order one, $\delta\omega(z)$ is of $1/N^2$: 
$\delta\omega=N^{-2} \tilde\omega_1 + \cO(N^{-3})$. 
Therefore at the leading order of the $1/N$ expansion this reduces to  
\be 
\hat {\cal K} \tilde\omega_{1}(z) = {d \over dT_z} \omega_{0,\half}(z),
\label{genus1loop2}
\ee
where we define $\hat {\cal K}$ by 
\be 
\hat {\cal K} f(z) :=\oint_{\cC_{\mbf R_+}}\!\!\! { dw \over 2\pi i} {W'(w)  \over z - w} f(w) - 2\omega_{0,\half}(z)f(z).
\ee
This equation is of the same form as the one without the hole correction by replacing $\omega_0, W_0$ with $\omega_{0,\half}, W$, respectively.
Since the above argument to solve this equation holds as it is by performing the replacement, a solution of \eqref{genus1loop2} is given by the same form as \eqref{resolvent0g1} where $a_i$ and $M_i^{(n)}$ are replaced with the ones including the hole correction. 
We emphasize that this simplification happens only at the genus one order and at higher order one may need to solve \eqref{loopg} in general. 

\section{Applications} 
\label{application}

In this section we apply the presented formulation developed in the previous section to a few examples. Firstly we apply to the three sphere partition function in $U(N)_k$ pure Chern-Simons theory in order to test the presented framework by comparing the exact result known for pure Chern-Simons theory as reviewed in \S\ref{pureCS}. 
Secondly we apply to $\cN=2$ $U(N)_k$ Chern-Simons theory with $n_F$ fundamental chiral multiplets and $\bar n_F$ anti-fundamental ones. 
This system does not admit the Fermi gas analysis in general and there may be no systematic way to study the system beyond the spherical limit except our formulation at present. 

\subsection{Pure Chern-Simons theory} 
\label{application2pureCS}

The matrix model potential for pure Chern-Simons theory is given by \eqref{defCSV2} with \eqref{pureCSparam}. 
For simplicity we first study a case such that there is no hole correction. 
Then $W'(w)=W_0'(w) = {\log w\over w\wt\lambda} + {1\over w}$.   

\begin{figure}
  \begin{center}
  \subfigure[]{\includegraphics[scale=.5]{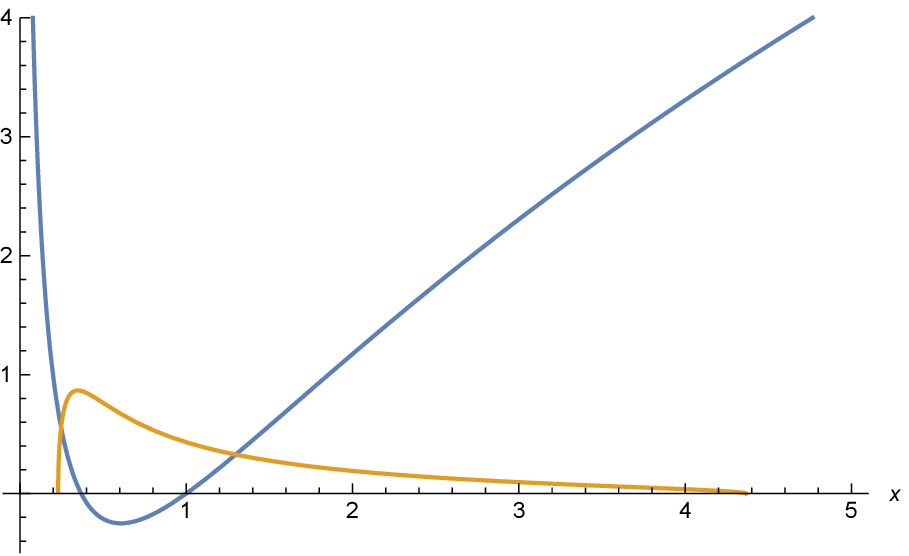}
\label{potden}
  }
  \qquad\qquad
  \subfigure[]{\includegraphics[scale=.5]{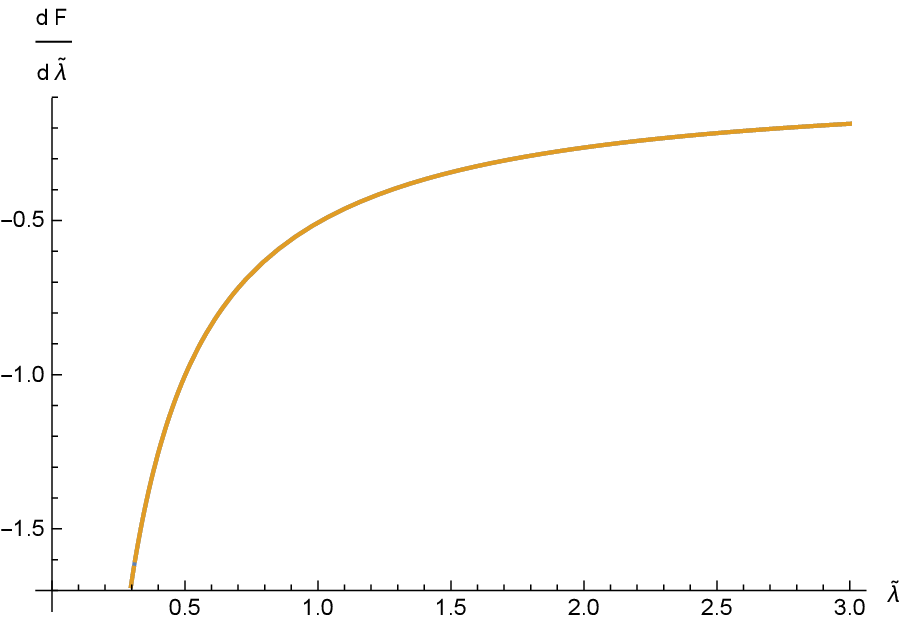}
\label{FE}
  }
  \end{center}
  \vspace{-0.5cm}
  \caption{In Fig.\ref{potden}, the blue and yellow curves depict the matrix model potential and the planar density function, respectively, when $\wt\lambda=0.5$. Eigenvalues tend to clump around the potential minimum. In Fig.\ref{FE}, the differentiation of the planar free energy with respect to $\wt\lambda$ is plotted. The blue curve depicts the result obtained by the resolvent method and the yellow one from the past exact result. They almost coincide.  }
  \label{}
\end{figure}

Let us first determine the planar resolvent. 
For $\wt\lambda>0$, the potential has only one stable minimum so we have only to consider a solution with one cut: $\supp(\rho_0)=[a_-,a_+]$ with $0<a_-<a_+$. 
In order to simplify the integration we start with the solution of the form \eqref{resolvent0another} with $k=1$: 
\beal{
\omega_0(z) =& {-h(z) \over 2z} \oint_{\cC_{[a_-,a_+]}}\!\! {dw \over 2\pi i} {{\log w\over \wt\lambda} + 1 \over (w - z)h(w)}
}
where $h(z) = \sqrt{(z-a_-)(z-a_+)}$. 
Inflating the contour we can compute the right-hand side as  
\beal{
\omega_0(z) =& {h(z) \over 2z} \oint_{\cC_{(-\infty,0]}}\!\! {dw \over 2\pi i} {{\log w\over \wt\lambda} \over (w - z)h(w)} +{h(z) \over 2z} {{\log z\over \wt\lambda} + 1 \over h(z)} \nn
=&{\log {(a_-+a_+) z - 2a_-a_+- 2 \sqrt{a_- a_+} h(z) \over  \left(-a_- -a_+  -2 h(z) +2 z\right)} \over 2\wt\lambda z} + {1 \over 2z}, 
\label{resolvent0pureCS}
}
where we computed the first term as 
\beal{
{h(z) \over 2\wt\lambda z} \int_{-\infty}^0\!\! {dw \over 2\pi i} {{(\log |w| - \pi i)-(\log |w| + \pi i) } \over (w - z)h(w)}
={h(z) \over 2\wt\lambda z}{\log {(a_-+a_+) z - 2a_-a_+- 2 \sqrt{a_- a_+} h(z) \over z \left(-a_- -a_+ -2 h(z)  +2 z\right)} \over h(z)}. 
}
The edges of the cut $a_-, a_+$ are determined by the asymptotic behavior around the infinity and the regularity around the origin. 
$\omega_0(z)$ can approach to ${1\over z}$ when $z \to -\infty$, which is achieved if and only if 
$
\log {a_-+a_+ +2 \sqrt{a_-a_+}  \over 2 +2} = \wt\lambda. 
$
$\omega_0$ is regular at the origin if and only if
$
 \log {-4 a_-a_+ \over -a_- - 2 \sqrt{a_-a_+}-a_+ } = -\wt\lambda. 
$
These can be solved as 
$
a_\pm = (e^{\wt \lambda\over 2} \pm \sqrt{e^{\wt\lambda}-1})^2.  
$
Note that $a_- a_+ =1$. 
Then the planar resolvent can be simplified as 
\beal{
\omega_0(z) =&{1 \over \wt\lambda z}\log {z+ 1 + h(z) \over 2  }. 
\label{resolvent0final}
}
The planar density function is computed as 
\be 
\rho_0(x) = {\tan^{-1} {\sqrt{(x-a_-)(a_+ - x)} \over 1+x} \over \pi \wt\lambda x}. 
\ee
This solution matches the one given in \cite{Marino:2004eq}, where the solution is expressed in the original coordinates. 
We plot the density function as well as the potential in Fig.\ref{potden}. 
The planar free energy is given by \eqref{freeenergy0} and its $\wt\lambda$ derivative is \eqref{freeenergyLambda}:
\beal{ 
{\partial F_{0} \over \partial \wt\lambda}  =&  {1 \over 12}  - {1 \over 2\wt\lambda^2} \int _{a_-}^{a_+}\!\! dx \rho_0(x)(\log x)^2. 
\label{Flambda}
} 
We could not perform the integration in the right-hand side analytically, so instead we evaluated it by numerics. The numerical result is in good agreement with the past exact result \eqref{F0ii} as seen in Fig.\ref{FE}. 

Next we study the genus one correction. 
Now we consider the one cut solution, so the genus one correction of the resolvent is given by 
\be
\omega_1(z) = {1\over 16} (\chi_-^{(2)}(z) +\chi_+^{(2)}(z) )- {1\over 8}{1\over a_- -a_+}(\chi_-^{(1)}(z) -\chi_+^{(1)}(z)), 
\ee
where $\chi_\pm$ are defined by \eqref{chi}. 
$M_\pm^{(1)}, M_\pm^{(2)}$ are computed by \eqref{Mik}. 
As done in the computation of the resolvent, 
the integration can be simplified by using \eqref{condition1}. 
\beal{
M_\pm^{(1)} =&{1\over a_\pm}\oint_{\cC_{[a_-,a_+]}}\!\! {dw \over 2\pi i} {wW_0'(w) \over h(w)(w- a_\pm) } 
= {1\over \wt\lambda a_\pm}\oint_{\cC_{[a_-,a_+]}}\!\! {dw \over 2\pi i} {\log w \over h(w)(w- a_\pm) } 
+{1\over a_\pm}\oint_{\cC_{[a_-,a_+]}}\!\! {dw \over 2\pi i} {1 \over h(w)(w- a_\pm) }.
\notag 
}
By inflating the contour to the infinity, the first term is computed as 
\beal{
{-1\over \wt\lambda a_\pm}\int_{-\infty}^0 \!\!{dy} {-1 \over h(y)(y- a_\pm) } 
={-1\over \wt\lambda a_\pm}\(\frac{2 \sqrt{w-a_\mp}}{(a_\pm-a_\mp) \sqrt{w-a_\pm}} \)\bigg|^0_{-\infty}
=& {-2 \over \wt\lambda a_\pm \sqrt{a_\pm} (\sqrt a_\pm+ \sqrt a_\mp)},   \notag
}
and the second term vanishes. Therefore 
$
M_\pm^{(1)} = {-2 \over \wt\lambda a_\pm^{3/2} (\sqrt a_\pm+ \sqrt a_\mp)}. 
$
In the same way $M_\pm^{(2)}$ are computed as 
$
M_\pm^{(2)}=\frac{2 \left(5 \sqrt{a_\pm}+4 \sqrt{a_\mp}\right)}{3 a_\pm^{5/2}\wt\lambda \left(\sqrt{a_\pm}+\sqrt{a_\mp}\right)^2}. 
$

The genus one correction of the free energy is computed by using \eqref{freeenergyLambda}: 
\beal{ 
{\partial F_{1} \over \partial \wt\lambda}  =&  {1 \over 12}  - {1 \over 2\wt\lambda^2} \int _{a_-}^{a_+}\!\! dx \rho_1(x)(\log x)^2 
= {1 \over 12} - {1 \over \wt\lambda^2}\int_{-\infty}^0{dx }\omega_1(x) \log (-x). 
}
This time we could perform the integral analytically:
\beal{ 
{\partial F_{1} \over \partial \wt\lambda}  =& \frac{e^{\wt\lambda} (\wt\lambda-2)+\wt\lambda+2}{24 (e^{\wt\lambda}-1 ) \wt\lambda}
= \frac{\wt\lambda \coth \left(\frac{\wt\lambda}{2}\right)-2}{24 \wt\lambda}. 
}
This result is in precise agreement with the past exact result of \eqref{F1ii} with $\zeta=0$. 

Next we consider a case where the matrix model potential has a hole correction
by the ``FI parameter''%
\footnote{ 
The usage of this terminology can be justified by adding some auxiliary fields into pure Chern-Simons theory so that the theory has $\cN=2$ supersymmetry.  
}: $W'(w) = {\log w\over w\wt\lambda} + {1+{\wt\zeta \over N} \over w}$ with $\wt\zeta=i{\zeta \over 2}$, which is still integrable as shown in \S\ref{pureCS}. 
As discussed in \S\ref{hole} we solve the usual saddle point equation $\omega(x-i\epsilon)+\omega(x+i\epsilon) =W'(x)$, which is correct up to the hole order. 
To emphasize the difference from the previous computation we denote the cut with prime added so that $\supp(\rho_{0,\half})=[a_-',a_+']$. 
Then a solution of this saddle point equation, $\omega_{0,\half}(x)$, is given by 
\beal{
\omega_{0,\half}(z) =& {-\acute h(z) \over 2z} \oint_{\cC_{[a_-',a_+']}}\!\! {dw \over 2\pi i} {{\log w\over \wt\lambda} + 1 + {\wt\zeta \over N} \over (w - z)\acute h(w)}
}
where $\acute h(z) = \sqrt{(z-a_-')(z-a_+')}$. 
This can be computed in the same way as previously and we obtain%
\footnote{ 
The planar resolvent and the hole one are determined from this by 
\be 
\acute\omega_{0}(z) ={\log {(a_-'+a_+') z - 2a_-'a_+' - 2 \sqrt{a_-' a_+'} \acute h(z) \over  \left(-a_-' -a_+'  -2 \acute h(z) +2 z\right)} \over 2\wt\lambda z} + {1\over 2z}, \quad 
\omega_{\half}(z) = {\wt\zeta \over 2z}, 
\ee
which should be done before the edges of the cuts are determined.   
}
\beal{
\omega_{0,\half}(z) ={\log {(a_-'+a_+') z - 2a_-'a_+' - 2 \sqrt{a_-' a_+'} \acute h(z) \over  \left(-a_-' -a_+'  -2 \acute h(z) +2 z\right)} \over 2\wt\lambda z} + {1+ {\wt\zeta \over N} \over 2z}. 
}
The edges of the cut are also determined in the same way. 
The result is 
$ a_\pm' =c a_\pm$, where $c=e^{-{\wt \lambda}{\wt\zeta\over N} }$ and $a_\pm$ are the same as previously.  
By using this the resolvent up to the hole order is simplified as 
$\omega_{0,\half}(z) =c^{-1} \omega_0(\acute z)$, where $\acute z= c^{-1}z$. 
As argued in \S\ref{genus1} the genus one resolvent is given by 
\beal{ 
\tilde\omega_{1}(z) =& {1\over 16} (\acute\chi_-^{(2)}(z) +\acute\chi_+^{(2)}(z) )- {1\over 8}{1\over a'_- -a'_+}(\acute\chi_-^{(1)}(z) -\acute\chi_+^{(1)}(z)), 
}
where $\acute\chi_i^{(n)}(z)$ is given by \eqref{chi} with $a_i$ replaced into $a'_i$. By using $ a_\pm' =c a_\pm$ the genus one resolvent including the FI term can be written as $\tilde\omega_{1}(z) =c^{-2} \omega_1(\acute z)$. 
Finally we compute the differentiation of the total free energy with respect to $\wt\lambda$.  
\beal{
{\partial\acute F \over \partial \wt\lambda}  
=& {7N^2 - 1 \over 12}- {N^2 \over 2\wt\lambda^2} \int _{a_-'}^{a_+'}\!\! dx \rho_{0,\half}(x)(\log x)^2 - {1 \over 2\wt\lambda^2} \int _{a_-'}^{a_+'}\!\! dx \acute\rho_{1}(x)(\log x)^2  + \cdots
}
where the ellipsis represents the terms of order $N^{-3}$. 
Then the 2nd term is computed as 
\beal{ 
- {N^2 \over 2\wt\lambda^2} \int _{a_-'}^{a_+'}\!\! dx \rho_{0,\half}(x)(\log x)^2    
=- {N^2 \over 2\wt\lambda^2}(\int _{a_-}^{a_+}\!\! d\acute x \rho_{0}(\acute x)(\log \acute x)^2+ (\log c)^2).\notag 
} 
The 3rd term is 
\beal{
- {1 \over 2\wt\lambda^2} \int _{a_-'}^{a_+'}\!\! dx \acute\rho_{1}(x)(\log x)^2 
=- {1 \over \wt\lambda^2} \int _{-\infty}^{0}\!\! dx \tilde\omega_{1}(x)\log(-x) 
=- {c^{-1} \over \wt\lambda^2} \int _{-\infty}^{0}\!\! d\acute x \omega_{1}(\acute x)\log(-\acute x). \notag 
}
As a result we obtain 
\beal{
{\partial\acute F \over \partial \wt\lambda}  
=&{\partial F \over \partial \wt\lambda}- {N^2 \over 2\wt\lambda^2} (\log c)^2 + \cdots
={\partial F \over \partial \wt\lambda}- {\wt \zeta^2 \over 2} + \cdots. 
}
This is in perfect agreement with the past exact result of \eqref{F1ii}. 

\subsection{$\cN=2$ Chern-Simons theory with arbitrary numbers of fundamental and anti-fundamental chiral multiplets} 
\label{application2N2CSV}

As another example we consider the matrix model of $\cN=2$ Chern-Simons theory with $n_f$ fundamental chiral multiplets and $\bar n_f$ anti-fundamental ones with the canonical R-charge. 
Let us set 
\be 
n_f^{(\pm)} = n_f \pm \bar n_f.
\ee
Without losing generality, we can assume that $n_f \geq \bar n_f$. 
The matrix model potential of this system is given by combining \eqref{N2CSV}
and \eqref{N2CSQ} 
\be
W(\phi_s)=  {1 \over 2\wt\lambda} (\log \phi_s)^2 + (1+{\wt\zeta \over N})\log \phi_s +{\bar n_f \over N} \log {\sqrt{\phi_s} + 1/\sqrt{\phi_s} \over 2 }-{n_f^{(-)}  \over N} \ell({-i\log\phi_s \over 2\pi}+\half).
\ee
Its derivative is 
\beal{ 
W'(w)
=& {\log w\over w\wt\lambda} + {1+{\wt\zeta\over N}\over w} + {1-w\over 2w(1+w)} \frac1N (n_f^{(-)}{i\log w \over 2\pi}-n_f^{(+)}{1 \over 2} ) ,
}
where we used $\ell'(z) = -\pi z \cot(\pi z)$.
Thus it is clear that the matrix model potential takes values in the complex number for a general number of chiral multiplets.

In order to determine the number of the cuts in the resolvent by identifying that of the potential minimum, we regard the matrix model potential as an analytic function with all the parameters. 
When $n_f^{(-)}$ is pure imaginary and $\wt\lambda, \wt\zeta, n_f^{(+)}$ are real, the potential becomes real. We fix the number of the cuts of the resolvent in this situation. 

We consider the large $k,N$ limit holding its ratio and other parameters $\wt\zeta, n_f^{(\pm)}$ fixed. 
In this limit the potential has only one stable minimum as in the case of pure Chern-Simons theory so we have only to consider a solution with one cut: $\supp(\rho_0)=[\mf a_-,\mf a_+]$ with $0<\mf a_-<\mf a_+$. 

We compute the resolvent up to the hole correction by  \eqref{resolvent0another} with $k=1$: 
\beal{
\omega_{0,\half}(z) =& {-h(z) \over 2z} \oint_{\cC_{[\mf a_-,\mf a_+]}}\!\! {dw \over 2\pi i} {{\log w\over \wt\lambda} + 1 +{\wt\zeta\over N} + {1-w\over 2(1+w)} \frac1N (n_f^{(-)}{i\log w \over 2\pi}-n_f^{(+)}{1 \over 2} ) \over (w - z)h(w)},
}
where $h(z) = \sqrt{(z-\mf a_-)(z-\mf a_+)}$. 
Inflating the contour we can compute the right-hand side by picking up the pole as 
\beal{
\omega_{0,\half}(z) =& {h(z) \over 2z} \oint_{\cC_{(-\infty,0]}}\!\! {dw \over 2\pi i} {{\log w\over \wt\lambda} + {n_f^{(-)} \over N} {i\log w \over 4\pi} {1-w \over 1+w}\over (w - z)h(w)} + {h(z) \over 2z} { \frac1N (-n_f^{(+)}{1 \over 2} ) \over (-1 - z)h(-1)} \nn
& +{h(z) \over 2z} {{\log z\over \wt\lambda} + 1 +{\wt\zeta\over N}+ {1-z\over 2(1+z)} \frac1N (n_f^{(-)}{i\log z \over 2\pi}-n_f^{(+)}{1 \over 2} ) \over h(z)} ,
}
where the first term is the contribution of the logarithmic branch cut $(-\infty, 0]$, the second one is the one of the pole at $w=-1$ and the third one is at $w=z$. 
We compute integrations in a manner that 
\beal{
\oint_{\cC_{(-\infty,0]}}\!\! {dw \over 2\pi i} { \log w \over (w - z)h(w)} 
=&  \int_{-\infty}^0\!\! {dw} { -1\over (w - z)h(w)} = \frac{f(z) - \log (z)}{h(z)}, \\
\oint_{\cC_{(-\infty,0]}}\!\! {dw \over 2\pi i} {(1-w)  \log w \over (1+w) (w - z)h(w)} 
=&  \int_{-\infty}^0\!\! {dw} \cP{ -(1-w) \over (w+1) (w - z)h(w)} = \frac{F(z)-F(-1)+\frac{(z-1) \log (z)}{h(z)}}{z+1}, 
}
where 
\be 
f(z) = \log [{(\mf a_-+\mf a_+) z - 2\mf a_-\mf a_+- 2 \sqrt{\mf a_- \mf a_+} h(z) \over  -\mf a_- -\mf a_+  -2 h(z) +2 z } ], \quad F(z)\text{:=}\frac{(1-z) f(z) }{h(z)}.
\ee
Then the resolvent becomes
\beal{
\omega_{0,\half}(z) =&{1 \over 2 z} \bigg[ f(z)({1\over \wt\lambda} + {n_f^{(-)}  \over N}{i \over 4\pi} {1-z \over 1+z} ) + 1 +{\wt\zeta\over N}  \nn
&+ {1\over (z+1)}{1 \over N} \( - n_f^{(-)} {i \over 4\pi} {h(z)}\frac{2 f(-1) }{h(-1)}+ n_f^{(+)}{1 \over 2}  {h(z)  \over h(-1)}  - n_f^{(+)} {1 - z \over 4} \)\bigg], 
}

The edges of the cut $\mf a_-, \mf a_+$ are determined by the asymptotic behavior around the infinity and the regularity around the origin. 
$\omega_{0,\half}(z)$ can approach to ${1\over z}$ when $z \to -\infty$, which is achieved if and only if 
\be 
{1 \over 2 } \bigg[ \log {\mf a_-+\mf a_+ +2 \sqrt{\mf a_-\mf a_+}  \over 2 +2}({1\over \wt\lambda} - {n_f^{(-)}  \over N}{i \over 4\pi} ) + 1 + {1 \over N}(\wt\zeta+ n_f^{(-)} {i \over 4\pi}\frac{2 f(-1) }{h(-1)}+ n_f^{(+)}{1 \over 2}  {-1  \over h(-1)}  - n_f^{(+)} {- 1 \over 4}) \bigg] = 1.
\ee
$\omega_{0,\half}$ is regular at the origin if and only if
\be 
\bigg[\log {-4 \mf a_-\mf a_+ \over -\mf a_- - 2 \sqrt{\mf a_-\mf a_+}-\mf a_+ } ({1\over \wt\lambda} + {n_f^{(-)}  \over N}{i \over 4\pi} ) + 1 +{1 \over N}(\wt\zeta - n_f^{(-)} {i \over 4\pi} {h(0)}\frac{2 f(-1) }{h(-1)}+ n_f^{(+)}{1 \over 2}  {h(0)  \over h(-1)}  - n_f^{(+)} {1 \over 4}) \bigg] = 0.
\ee
From these equations the edges of the cut are determined order by order in $1/N$.

As argued in \S\ref{hole} the planar resolvent should be determined so as to have the same cut as that of $\omega_{0,\half}(z)$.
Since the leading part of the potential in the large $N$ limit is unchanged, the form of the planar resolvent is unchanged except the edges of the cut: \eqref{resolvent0pureCS} with $a_i \to \mf a_i$. 
The genus half resolvent is determined before the edges of the cut are expanded in the $1/N$ power series and given by
\be 
\omega_\half(z) ={1 \over 2 z} \bigg[{\wt\zeta } + f(z)( {n_f^{(-)}}{i \over 4\pi} {1-z \over 1+z} ) + {1\over (z+1)}( - n_f^{(-)} {i \over 4\pi} {h(z)}\frac{2 f(-1) }{h(-1)}+ n_f^{(+)}{1 \over 2}  {h(z)  \over h(-1)}  - n_f^{(+)} {1 - z \over 4})\bigg].
\ee
Then as argued in \S\ref{genus1}, 
the genus one resolvent is given by   
\beal{ 
\tilde\omega_{1}(z) =& {1\over 16} (\chi_-^{(2)}(z) +\chi_+^{(2)}(z) )- {1\over 8}{1\over  \mf a_- - \mf a_+}(\chi_-^{(1)}(z) -\chi_+^{(1)}(z)),
}
where $\chi_i^{(n)}(z)$ are given by \eqref{chi} with $a_\pm$ replaced into $\mf a_\pm$.  
The $\wt\lambda$ derivative of the free energy up to the genus one order is given by
\beal{
{\partial F \over \partial \wt\lambda}  
=&- {\partial \log \mf N \over \partial \wt\lambda}   - {N^2 \over 2\wt\lambda^2} \int _{\mf a_-}^{\mf a_+}\!\! dx \rho_{0,\half}(x)(\log x)^2  - {1 \over \wt\lambda^2}\int_{-\infty}^0{dx }\tilde\omega_1(x) \log (-x).
}

It is known that this system has the dual description known as Seiberg-like duality \cite{Benini:2011mf}. 
The dual theory is U$(N')_{-k}$ Chern-Simons theory with $n_f$ fundamental and $\bar n_f$ anti-fundamental chiral multiplets with $n_f \bar n_f$ mesonic operators as well as some monopole operators with a suitable superpotential, where $N'$ depends generally on $N,k,n_f,\bar n_f$, which is still in the class investigated in this paper. 
It would be interesting to test the duality from our general solution. We hope to come back to this problem in a future publication. 

\section{Discussion}
\label{discussion} 

In this paper we have performed a general analysis on a class of matrix models describing Chern-Simons matter theories on three sphere incorporating the standard technique of the $1/N$ expansion developed in the study of ordinary hermitian matrix models. 
We have derived the loop equation for all orders in the $1/N$ expansion and presented its explicit solution up to the genus one order when the potential has $1/N$ correction. 
We have applied the formulation to pure Chern-Simons theory and confirmed that the presented solution reproduces the exact result known in the past. 
We have also applied the framework to $\cN=2$ Chern-Simons theory with arbitrary numbers of fundamental and anti-fundamental chiral multiplets and obtained a formal expression of the solution up to the genus one order in the $1/N$ expansion.

This paper mainly focused on construction of the framework to solve a class of matrix models. 
We are very much interested in applying the formula obtained in this paper to a duality pair of Chern-Simons matter systems and testing the bosonization duality holds at the next leading order in the $1/N$ expansion.
In particular it would be interesting to develop the presented large $N$ technique in a class of unitary matrix models which arises as partition function of Chern-Simons matter theories on $\mbf S^2\times \mbf S^1$.
For a class of Chern-Simons vector models, 
the effective matrix model potential was determined exactly in the leading of the large $N$ limit \cite{Jain:2013py} and the three dimensional bosonization was confirmed at the order. 
We hope that the formulation developed in this paper is useful for future study in this direction. 

In this paper in order to study beyond the planar limit 
we adopted the iterative procedure given in \cite{Ambjorn:1992gw,Akemann:1996zr}. 
Another iterative approach has been proposed by using the Feynman graph of the trivalent vertexes \cite{Eynard:2004mh,Chekhov:2006rq}. 
It would be interesting to reformulate the presented formula in this note in terms of the different approach. 

Another interesting question is whether this class of matrix models has the equivalent description of some two dimensional CFT as the ordinary hermitian matrix models \cite{Mironov:1990im,Dijkgraaf:1990rs,Fukuma:1990yk}. (See also \cite{Kostov:1999xi}.)
Naively this answer seems to be no due to the fact that the degrees of freedom in 3d system are much bigger than those of 2d one in a generic situation.  
However we have a suspicion that the answer could be yes for a certain matrix model of this kind, intuitively because vector models coupling to Chern-Simons theory appear as effective field theory of anyonic system \cite{Zhang:1988wy,Fradkin:1991wy} and wave function describing a quantum Hall state known as Laughlin wave function \cite{Laughlin:1983fy} is given by a correlator of certain 2d (rational) CFT \cite{Moore:1991ks}. 
In fact it was shown that this answer becomes yes for a similar class of matrix models to the one studied in this paper \cite{Nedelin:2015mio}, where the corresponding CFT is identified with $q$-deformed one. 
Exploring this question is left to future work.

There is a straight-forward generalization of the presented formulation to a different gauge group \cite{Halmagyi:2003fy} or two matrices. 
This generalization to two matrices is important for the application to higher supersymmetric Chern-Simons matter theories such as the ABJM theory \cite{Aharony:2008ug}.  
The $1/N$ correction of the free energy in the ABJM theory was computed in \cite{Fuji:2011km,Marino:2011eh,Mezei:2013gqa}. In this development a new technique called Fermi gas approach was invented \cite{Marino:2011eh}. This approach is powerful to study non-perturbative aspects of the ABJM theory from the $\mbf S^3$ partition function \cite{Hatsuda:2012dt,Hatsuda:2013oxa}. 
It is an important problem to test whether the traditional techniques of matrix models can reproduce the results obtained by new ones in recent development beyond the spherical limit. 

We hope to come back to these issues in the near future. 

\section*{Acknowledgments}

The author would like to thank S.~Sugimoto and T.~Takayanagi for valuable discussions and comments on the draft. 
The author would also like to thank Y.~Imamura for a helpful comment on the first version of this paper. 

\appendix 
\section{Partition function of pure Chern-Simons theory on $\mbf S^3$ }
\label{pureCS} 

In this appendix we briefly overview the three sphere partition function in $U(N)_k$ pure Chern-Simons theory and a derivation of its large $N$ expansion used in the main text. 

The partition function is defined formally by a path integral over the gauge field on $\mbf S^3$ such that%
\footnote{  
Here $k$ is the renormalized Chern-Simons coupling constant so that $k=\kappa +\sgn(\kappa)N$, where $\kappa$ is the level of the corresponding WZW model. 
}
\be 
Z_{\rm CS} = \int \cD A e^{-{i k\over 2\pi} \int_{\mbf S^3} (\half A \wedge dA - {i \over 3} A\wedge A \wedge A) }.
\label{defpureCSpf} 
\ee
The classic paper \cite{Witten:1988hf} demonstrated explicitly in the case of SU(2) that this can be exactly determined as a function of the Chern-Simons level without performing the path integral by clarifying its relation to a modular transformation matrix of the characters in the corresponding Affine Lie algebra. 
Generalization to arbitrary gauge group is straight-forward. 
Since modular transformation matrices were already determined in general Affine Lie algebras \cite{Kac:1984mq}, 
the exact result of the partition function for $U(N)_k$ pure Chern-Simons theory was given by
\be 
Z_{\rm CS}= k^{-{N \over 2}} \prod_{I=1}^{N-1} (2\sin{\pi I \over k})^{N-I}.
\label{pureCSpf}
\ee

After this exact result was studied in terms of the $1/N$ and $1/k$ expansions \cite{Camperi:1990dk,Periwal:1993yu},
it was insightfully observed that the Chern-Simons partition function \eqref{pureCSpf} exactly matches that of the topological string theory on a Calabi-Yau three-fold background by identifying the string coupling constant with the pure imaginary Chern-Simons level \cite{Gopakumar:1998ii}. 
This lead to the conjecture of gauge/geometry duality \cite{Maldacena:1997re} between Chern-Simons/Topological string theories \cite{Gopakumar:1998ki,Gopakumar:1998jq}.

Subsequently it was pointed out that the partition function \eqref{defpureCSpf} 
reduces to a matrix model such that \cite{Marino:2002fk}
\beal{
Z_{\rm CS} =&{(-)^{N(N-1) \over 2} e^{-\pi (N-1)N(N+1) \over 6 i k}i^{N^2 \over 2}\over (2\pi)^N N!}  \int_{\mbf R^N}\!\! d^N\!{\sigma} e^{-i {k \over 4\pi}\sum_{s=1}^N \sigma_s^2} { \prod_{t \not= s}^N 2\sinh({\sigma_s-\sigma_t\over2}) }. 
\label{pureCSpfMM}
}
This matrix model was extensively studied in relation to the topological string theory \cite{Aganagic:2002qg,Aganagic:2002wv}. 
With the help of the Weyl denominator formula the matrix integral was explicitly performed by gaussian integration in perfect agreement with \eqref{pureCSpf} \cite{Kapustin:2009kz}.
This matrix model was also evaluated exactly by the orthogonal (or characteristic) polynomial method in accordance with \eqref{pureCSpf} \cite{Tierz:2002jj}. The orthogonal polynomials associated with this matrix model were found to be Stiltjes-Wigert polynomials. 

Let us compute the matrix model \eqref{pureCSpfMM} in a Fermi gas like approach  \cite{Marino:2011eh} including the ``FI term'':
\beal{\!
Z_{\rm CS} =&{(-)^{N(N-1) \over 2} e^{-\pi (N-1)N(N+1) \over 6 i k}i^{N^2 \over 2}\over (2\pi)^N N!}  \int_{\mbf R^N}\!\! d^N\!{\sigma} e^{-i {k \over 4\pi}\sum_{s=1}^N \sigma_s^2 -i \half \zeta\sum_{s=1}^N \sigma_s } { \prod_{t \not= s}^N 2\sinh({\sigma_s-\sigma_t\over2}) }. 
\label{pureCSpfMM}
}
For this purpose we rewrite the partition function as a determinant by using the Weyl denominator formula 
\be 
\prod_{s>t}2\sinh{\sigma_s - \sigma_t \over 2} =\underset{s,t}\det[ e^{\sigma_s(t-{N+1 \over 2})} ].
\label{Weylden} 
\ee
From this formula we can show that  
\beal{
\prod_{s\not=t}[2\sinh{\sigma_s - \sigma_t \over 2} ]= N! \determinant{s,t} [e^{\sigma_s(-s +t)}],
}
which enables us to rewrite the partition function as 
\beal{ 
Z_{CS} =& {(-)^{N(N-1)\over2}e^{-\pi (N-1)N(N+1) \over 6 i k}i^{N^2 \over 2}} \determinant{s,t} \bigg[\int  {d\sigma_s \over 2\pi}e^{-(i {k \over 4\pi}\sigma_s^2 + i\half \zeta \sigma_s) } e^{\sigma_s(-s+t)}\bigg]. 
}
The inside of the determinant is computed by gaussian integration as follows. 
\beal{
\int  {d\sigma_s \over 2\pi}e^{-(i {k \over 4\pi}\sigma_s^2 + i\half \zeta \sigma_s) } e^{\sigma_s(-s+t)}
=\sqrt{1\over{ik}} e^{-\pi i (-i\half\zeta-s+t)^2 \over k}. 
}
Plugging this back gives 
\beal{ 
Z_{CS} 
=& {(-)^{N(N-1)\over2}e^{-\pi (N-1)N(N+1) \over 6 i k}i^{N^2 \over 2}} \determinant{s,t} \bigg[\sqrt{1\over{ik}} e^{- \pi i (-i\half\zeta-s+t)^2 \over k} \bigg]
= k^{-N\over2}e^{\pi iN \zeta^2 \over 4k }\prod_{s>t}2\sin[{\pi (s-t) \over k}]  
\notag 
}
where in the 2nd equation we used $\determinant{s,t}[f_s M_{s,t} ]=(\prod_{s}f_s)\determinant{s,t}[M_{s,t} ]$ and the Weyl denominator formula \eqref{Weylden}. 
By using the formula $\prod_{s>t}2\sin[{\pi (s-t) \over k}]  = \prod_{I=1}^N (2\sin{\pi I \over k})^{N-I}$
we obtain 
\be
Z_{CS} = k^{-{N \over 2}} e^{\pi iN \zeta^2 \over 4k } \prod_{I=1}^{N-1} (2\sin{\pi I \over k})^{N-I}.
\ee
Then the free energy is computed as 
\be 
F_{CS} = - \log Z_{CS} =  {N\over 2} \log k -{\pi iN \zeta^2 \over 4k } - \sum_{I=1}^{N-1} (N-I) \log  (2\sin{\pi I \over k}).
\label{pureCSFE}
\ee

The $1/N$ expansion was done as follows \cite{Gopakumar:1998ii,Gopakumar:1998ki}. 
The expansion coefficients are determined as functions of $\lambda = {N \over k}$. 
\be 
F_{\rm CS} = \sum_{g=0}^\infty N^{2 - 2g } F_{g}(\lambda). 
\label{pureCSFE1byN}
\ee
By using 
\be 
\sin x = x \prod_{n=1}^\infty (1 - ({x \over \pi n})^2), \quad 
\log (1 -x) = - \sum_{m=1}^\infty {x^m\over m}, 
\label{formulasin}
\ee
the second term in \eqref{pureCSFE} can be expanded as 
\beal{ 
&- \sum_{I=1}^{N-1} (N-I) \log  (2\sin{\pi I \over k})
=\sum_{I=1}^{N-1} (I - N) \log  (2 {\pi I \lambda \over N} \prod_{n=1}^\infty (1 - ({\pi I \lambda /N \over \pi n})^2) ) \nn
=&-{N(N-1) \over 2}  \log  {2\pi\lambda \over N}+\sum_{I=1}^{N-1} (I - N) \log  I+\sum_{m=1}^\infty {\zeta(2m) \over m} ({\lambda \over N })^{2m} \sum_{I=1}^{N-1} ( N - I) I^{2m}, 
\label{secondterm}
}
where $\zeta(m)$ is the zeta function defined by $\zeta(s) := \sum_{n=1}^\infty {1\over n^s}$.
The second term in \eqref{secondterm} can be expressed by using the Barn's function as 
$\sum_{I=1}^{N-1} (I - N) \log  I= - \log G(N+1)$, whose large $N$ expansion is known:  
\beal{ 
\log G(N+1)=& N^2(\half\log N - {3\over 4}) + {N \over 2} \log 2\pi - {B_2 \over 2}\log N+\zeta'(-1)+ \sum_{g=2}^\infty {B_{2g} \over 2g(2g-2) }N^{2 - 2g}.\notag
}
Here $B_n$ is the $n$-th Bernoulli number defined by%
\footnote{ 
Our definition of the Bernoulli number is different from the one adopted in several past literatures such as \cite{Periwal:1993yu,Gopakumar:1998ii,Gopakumar:1998ki}. The difference is $B_g^{(\rm there)} =(-)^{g-1} B_{2g}^{(\rm here)}$. 
}
\be 
{x \over e^x -1} = \sum_{n=0}^\infty {B_n \over n!} x^n.
\ee
The summation in the last term in \eqref{secondterm} can be done by using a formula such that%
\footnote{ 
This can be proven by using 
\beal{ 
\sum_{I=1}^N I^{m} =& {N^{m+1} \over m+1} + {N^{m} \over 2} + \sum_{g=1}^{[{m \over 2}]} { B_{2g} \over 2g}
\begin{pmatrix}
 m \\
 2g-1\\ 
\end{pmatrix} 
N^{m - 2g +1},
}
where $[x]$ is the integer part of $x$. 
}
\be 
\sum_{I=1}^{N-1} ( N - I) I^{2m}
= {N^{2m+2} \over (2m+1)(2m+2) } + \sum_{g=1}^m 
\begin{pmatrix}
 2m \\
 2g-2 
\end{pmatrix}
{- B_{2g} \over 2g }N^{2m+2-2g}.
\ee
Plugging these back, we obtain the free energy as
\beal{  
F_{\rm CS} =& {N\over 2} \log k -{N(N-1) \over 2}  \log  {2\pi\lambda \over N}+\sum_{m=1}^\infty {\zeta(2m) \over m} {\lambda}^{2m} \bigg[ {N^{2} \over (2m+1)(2m+2) } + \sum_{g=1}^m 
\begin{pmatrix}
 2m \\
 2g-2 
\end{pmatrix}
{- B_{2g} \over 2g }N^{2-2g}   \bigg]  \nn
&-{\pi iN \zeta^2 \over 4k } -\bigg(
N^2(\half \log N - {3\over 4})+ {N \over 2} \log 2\pi - {B_2 \over 2}\log N+\zeta'(-1) +\sum_{g=2}^\infty {B_{2g} \over 2g(2g-2) }N^{2 - 2g}\bigg) \nn
=&N^2 \bigg[ -\half \log 2\pi\lambda + {3\over 4} + \sum_{m=1}^\infty {\zeta(2m) \over m} {\lambda^{2m} \over (2m+1)(2m+2) } \bigg]+ {B_2 \over 2}\log N -{\pi iN \zeta^2 \over 4k } \nn
&-\zeta'(-1)  +\sum_{m=1}^\infty {\zeta(2m) } {\lambda}^{2m}  
{- B_2 \over 2m } +\sum_{g=2}^\infty N^{2 - 2g} {- B_{2g} \over 2g(2g-2) }\bigg[1 +\sum_{m=g}^\infty {\zeta(2m) } {\lambda}^{2m} 2 
\begin{pmatrix}
 2m-1 \\
 2g-3 
\end{pmatrix}
\bigg]. \notag
}
As a result, the coefficients in the $1/N$ expansion of the form \eqref{pureCSFE1byN} are determined as 
\beal{
F_{0} =& -\half \log 2\pi\lambda + {3\over 4} + \sum_{m=1}^\infty {\zeta(2m) \over m} {\lambda^{2m} \over (2m+1)(2m+2) }, \label{F0} \\ 
F_{1} =&-\zeta'(-1) -{\pi i\lambda \zeta^2 \over 4 }+ \sum_{m=1}^\infty {\zeta(2m) } {\lambda}^{2m}  
{-B_2 \over 2m }, \label{F1} \\ 
F_{g} =&{- B_{2g} \over 2g(2g-2) }\bigg[1 +\sum_{m=g}^\infty {\zeta(2m) } {\lambda}^{2m} 2 
\begin{pmatrix}
 2m-1 \\
 2g-3 
\end{pmatrix}
\bigg], \qquad (g\geq2). \label{Fg}
}

A few comments are in order. The leading term in the $1/N$ expansion, $F_0$, can be obtained directly from \eqref{pureCSFE} by taking the large $N$ limit \cite{Camperi:1990dk}. 
\beal{ 
F_0=& \lim_{N\to\infty}{F_{\rm CS} \over N^2} = \lim_{N\to\infty} - {1\over N} \sum_{I=1}^{N-1} (1-{I \over N}) \log  (2\sin{\pi \lambda I \over N})
=- \int_0^1 d\tau (1-\tau ) \log  (2\sin{\pi \lambda \tau}), \notag
}
where in the last equation we used the definition of the Riemann integral. 
This can be further computed by using \eqref{formulasin} as 
\beal{
F_0=& {\pi i \over 4} -{1\over 6} i \pi\lambda + { \zeta(2) \over 2i\pi \lambda} + {\zeta(3) \over (2\pi\lambda)^2} -{1 \over (2\pi\lambda)^2} \Li_3(e^{-2\pi i\lambda}), 
\label{F0ii} 
}
where $\Li_s(z)$ is the polylogarithm defined by $\Li_s(z) = \sum_{n=1}^\infty {z^n \over n^s}$. 
The next leading term $F_1$ can be simplified by using \eqref{formulasin} as follows. 
\beal{
F_{1}=&-\zeta'(-1) -{\pi i\lambda \zeta^2 \over 4 } + {1\over 12} \log {\sin\pi\lambda \over \pi\lambda}.  
\label{F1ii}
}

\bibliographystyle{utphys}
\bibliography{CSpf}

\end{document}